\documentclass[aps,prc,twocolumn,superscriptaddress]{revtex4-1}
\usepackage{url}                         % URL format
\usepackage{subfigure}
\usepackage{epsfig}
\usepackage{amsmath,amssymb}
\usepackage{longtable}
\usepackage{lineno}
\usepackage{soul}
\usepackage{color}

\renewcommand{\arraystretch}{2}

\begin{document}

%\preprint{}


\title{The n$^3$He Experiment: Parity Violation in Polarized Neutron Capture on $^{3}$He}

%\thanks{}
\author{M.~McCrea}
\affiliation{University of Winnipeg}
\author{M.~L.~Kabir }
\affiliation{University of Kentucky}
\author{N.~Birge}
\affiliation{University of Tennessee}
\author{C.~E.~Coppola}
\affiliation{University of Tennessee}
\author{C.~Hayes}
\affiliation{University of Tennessee}
\author{E.~Plemons}
\affiliation{University of Tennessee}
\author{A.~Ram\'irez-Morales}
\affiliation{Universidad Nacional Aut\'{o}noma de M\'{e}xico}
\author{E.~M.~Scott}
\affiliation{University of Tennessee}
\author{J.~Watts}
\affiliation{University of Tennessee Chattanooga}
%\author{R.~Alarcon}
%\affiliation{Arizona State University}
\author{S.~Baessler}
\affiliation{University of Virginia}
\affiliation{Oak Ridge National Laboratory}
\author{L.~Barr\'on-Palos}
\affiliation{Universidad Nacional Aut\'{o}noma de M\'{e}xico}
\author{J.~D.~Bowman}
\affiliation{Oak Ridge National Laboratory}
\author{C.~Britton Jr}
\affiliation{Oak Ridge National Laboratory}
\author{J.~Calarco}
\affiliation{University of New Hampshire}
\author{V.~Cianciolo}
\affiliation{Oak Ridge National Laboratory}
\author{C.~B.~Crawford}
\affiliation{University of Kentucky}
\author{D.~Ezell}
\affiliation{Oak Ridge National Laboratory}
\author{N.~Fomin}
\affiliation{University of Tennessee}
\author{I.~Garishvili}
\affiliation{University of Tennessee}
\author{M.~T.~Gericke}
\affiliation{University of Manitoba}
\email[]{Michael.Gericke@umanitoba.ca}
\author{G.~L.~Greene}
\affiliation{University of Tennessee}
\affiliation{Oak Ridge National Laboratory}
\author{G.~M.~Hale}
\affiliation{Los Alamos National Laboratory}
\author{J.~Hamblen}
\affiliation{University of Tennessee Chattanooga}
\author{E.~Iverson}
\affiliation{Oak Ridge National Laboratory}
\author{P.~E.~Mueller}
\affiliation{Oak Ridge National Laboratory}
\author{I.~Novikov}
\affiliation{Western Kentucky University}
\author{S.~Penttila}
\affiliation{Oak Ridge National Laboratory}
\author{C.~Wickersham}
\affiliation{University of Tennessee Chattanooga}

\collaboration{The n3He Collaboration}
\noaffiliation

\date{\today}

\begin{abstract}
Significant progress has been made to experimentally determine a complete set of the parity-violating (PV) weak-interaction amplitudes between nucleons. In this paper we describe the design, construction and operation of the n$^3$He experiment that was used to measure the PV asymmetry $A_{\mathrm{PV}}$ in the direction of proton emission in the reaction $\vec{\mathrm{n}} + {^3}\mathrm{He} \rightarrow {^3}\mathrm{H} + \mathrm{p}$, using the capture of polarized cold neutrons in an unpolarized gaseous $^3\mathrm{He}$ target. This asymmetry has was recently calculated~\cite{Viviani,Viviani2}, both in the traditional style meson exchange picture, and in effective field theory (EFT), including two-pion exchange. The high precision result (published separately) obtained with the experiment described herein forms an important benchmark for hadronic PV (HPV) theory in few-body systems, where precise calculations are possible. To this day, HPV is still one of the most poorly understood aspects of the electro-weak theory. The calculations estimate the size of the asymmetry to be in the range of $(-9.4 \rightarrow 3.5)\times 10^{-8}$, depending on the framework or model. The small size of the asymmetry and the small overall goal uncertainty of the experiment of $\delta A_{\mathrm{PV}} \simeq 1\times10^{-8}$ places strict requirements on the experiment, especially on the design of the target-detector chamber. In this paper we describe the experimental setup and the measurement methodology as well as the detailed design of the chamber, including results of Garfield++ and Geant4 simulations that form the basis of the chamber design and analysis. We also show data from commissioning and production and define the systematic errors that the chamber contributes to the measured $A_{\mathrm{PV}}$. We give the final uncertainty on the measurement.

\end{abstract}

%\pacs{}
%\keywords{}

\maketitle

\section{Introduction}
\label{sect:intro}
In a strangeness conserving weak interaction between a pair of nucleons isospin can change by $\Delta I$ = 0, 1, or 2 ~\cite{Adel,Musolf,Haxton}. Desplanques, Donoghue, and Holstein (DDH) introduced some time ago a meson exchange model to describe the hadronic weak interaction (HWI)~\cite{DDH}. In DDH the HWI is parametrized by parity-violating (PV) couplings of the light mesons ($\pi,\rho,\omega$) at one vertex and the strong coupling of the mesons at the other vertex, leading to 7 weak coupling constants corresponding to different meson-isospin combinations. Then PV observables are expressed as a sum of the couplings with weights defined by the strong interaction, calculated from the latest NN potential. The DDH model describes the HWI up to the pion production threshold. In the effective field theory (EFT), the HWI is modeled by writing down a Lagrangian in terms of nucleon degrees of freedom, which is consistent with the symmetries of QCD, and then performing a perturbative expansion in small parameter $(q/\Lambda_\chi)$ ~\cite{Schindler,Haxton,Musolf,Erler,Liu}, where $q$ is the typical internal momentum transfer and $\Lambda \sim 1$ GeV is the strong QCD scale. To lowest order, one obtains pionless EFT, in which the resulting PV NN potential contains 5 low energy coupling constants. However, in reality, the relative size of the NN weak amplitudes in different spin and isospin channels might not be determined entirely by simple symmetry arguments, since they are sensitive to the short-range correlations between confined quarks in the nucleon and to strong NN correlations. The NN weak amplitudes are therefore a good probe of the poorly understood confinement and chiral symmetry breaking dynamics of QCD.

New efforts to calculate parity violating NN observables, starting from ($\chi$PT based) EFT are being undertaken, but accurate experimental results are needed to relate them to modern NN potentials such as AV18. The first approach to calculate the weak nucleon coupling amplitude for the $\Delta I=1$ channel on the lattice has been carried out as well~\cite{Wasem} and new, improved LQCD calculations are underway. The experimental challenge lies in finding feasible few-body experiments, which can measure with high accuracy observables sensitive to different isospin channels that, together, over-constrain the calculated weak couplings. This has been possible in the past few years, due to the availability of new, high intensity neutron facilities, such as the Spallation Neutron Source (SNS) at Oak Ridge National Laboratory (ORNL).

The n$^3$He experiment was the most recent precision test of the HWI, measuring the PV asymmetry $A_{\mathrm{PV}}$, in the correlation between the spin of incoming neutrons $\hat{s}_n$ and the outgoing momentum of protons $(\vec{k}_{\mathrm{p}})$ in the reaction $\vec{\mathrm{n}} + ~^3\mathrm{He} \longrightarrow ~^3\mathrm{H} + \mathrm{p} + 765 ~\rm{keV}$. The size of the asymmetry has been calculated both based on the DDH formalism ($A_{\mathrm{PV}} = (-9.4 \rightarrow 2.5) \times 10^{-8}$) ~\cite{Viviani,Girlanda} and the EFT framework ($A_{\mathrm{PV}} = 1.7\times10^{-8}$ for $\Lambda_\chi = 500~\rm{MeV}$  and $A_{\mathrm{PV}} = 3.5\times10^{-8}$ for $\Lambda_\chi = 600~\rm{MeV}$~\cite{Viviani2}). The n$^3$He experiment is an important independent probe of the hadronic weak couplings. Together with NPDGamma ~\cite{Gericke1,Blyth:2018aon}
%citation added to bibliograph as Blyth:2018aon by M.McCrea orignal citer can check if correct.
%((D. Blyth et al. (NPDGamma Collaboration), arXiv:1807.10192 (2018).)),
neutron spin rotation~\cite{Snow},
data from p-p scattering ~\cite{Evershiem,Kistryn,Balzer},
and the zero measurement of the absolute value of circular polarization from unpolarized $^{18}$F nuclei that depends only on the $\Delta I=1$ terms in the HWI, the n$^3$He experiment provides the 4th measurement in a few body system with straightforward theoretical interpretation of the results, constraining the full set of weak nucleon-nucleon couplings.

\section{Overview of the n$^3$He experiment} \label{sect:exptsetup}

\subsection{The Fundamental Neutron Physics Beamline} \label{sect:beamline}
\begin{figure}[htb]
  \begin{center}
  \includegraphics[width=0.9\columnwidth]{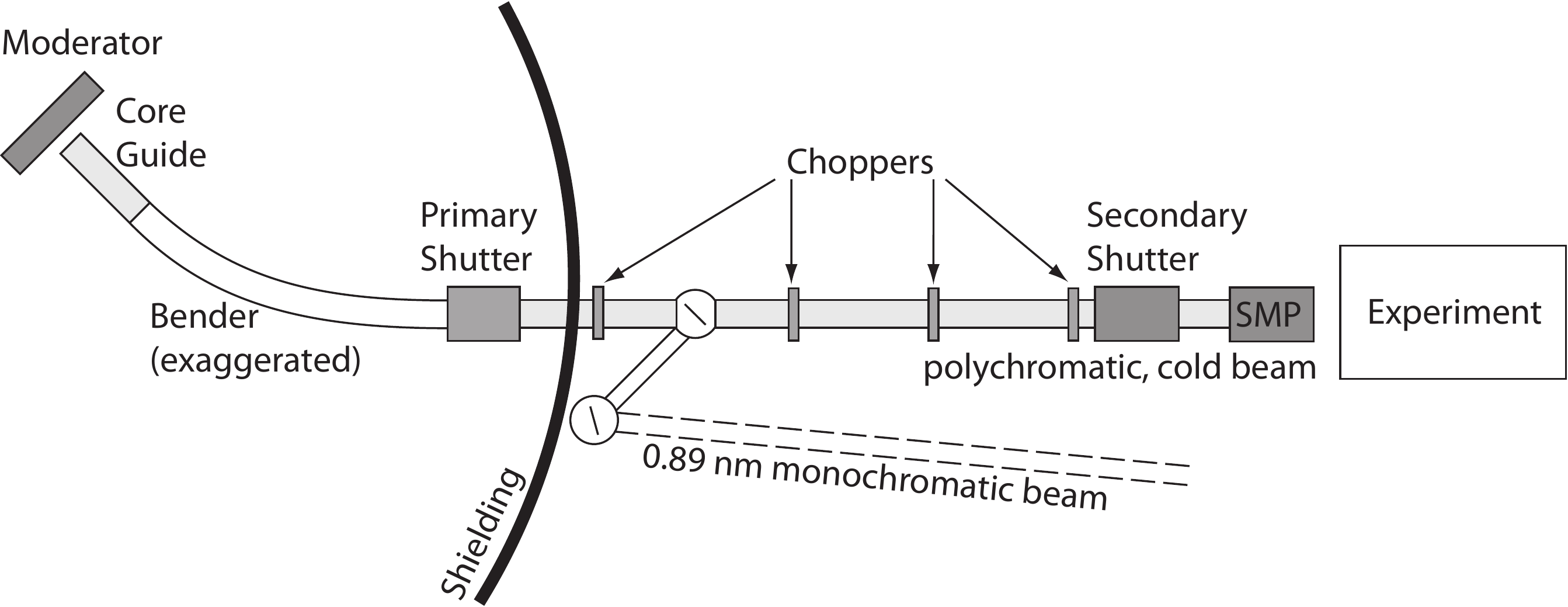}
    \caption{Schematic of the FnPB at the SNS. Neutrons exiting the liquid hydrogen moderator are guided by reflections from the $m=3$ supermirror surfaces of the neutron guide ~\cite{Fomin}. A curved guide section removes high energy line-of-sight backgrounds. The two choppers after the moderator are used to select a specific time-of-flight frame in each neutron pulse. Fig.~\ref{fig:n3HeSetup} shows the n$^3$He experimental setup.}
    \label{fig:FnPBSchem}
  \end{center}
\end{figure}

The n$^3$He experiment ran at the Fundamental Neutron Physics Beam line (FnPB) at the Spallation Neutron Source at Oak Ridge National Laboratory, from December 2014 to December 2015. The beam characteristics at the FnPB are described in ~\cite{Fomin} and references within. A schematic of the FnPB is shown in Fig.~\ref{fig:FnPBSchem} and the n$^3$He setup is briefly described below.

\begin{figure}[htb]
  \begin{center}
  \includegraphics[width=0.8\columnwidth]{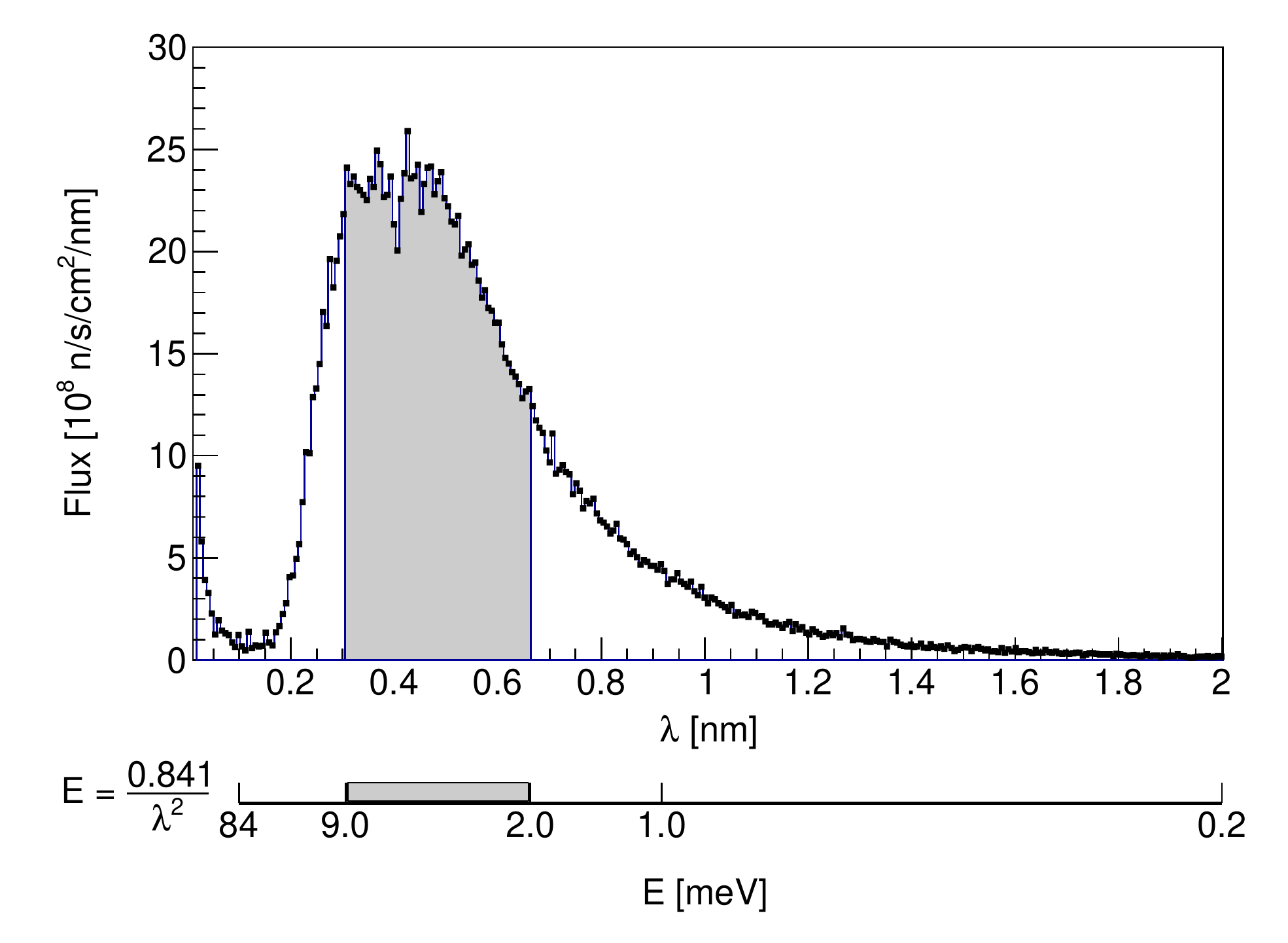}
  \caption{Measured neutron spectrum out of the moderator~\cite{Fomin}, scaled to 1.4 MW proton beam power, as a function of neutron wavelength and energy. The energy range selected for n$^3$He, using the beam choppers, is indicated in gray. In reality, the sharp edges of the gray area in the figure have a slight slope that is produced by the moving edges of the chopper.}
  \label{fig:SNSFullPulse}
  \end{center}
\end{figure}

Pulses of high energy spallation neutrons were produced by $\simeq 1\mu$~s long 1~GeV proton pulses interacting at $60~\,\mathrm{Hz}$ with a circulated liquid Mercury target. The MeV spallation neutrons are cooled by a liquid hydrogen moderator. This thermalization process produces pulses of cold neutrons, with a Maxwell-Boltzmann energy distribution related to the temperature of the moderator, see Fig.~\ref{fig:SNSFullPulse}. This allows for accurate time-of-flight (TOF) measurements of  the neutron energy. The energy information was a necessity for data analysis, effective neutron polarization on the target and for studies of sources of systematic effects. From the moderator, the neutrons were guided to the experiment by a supermirror neutron guide ~\cite{Fomin} with cross sectional area of $10\,\mathrm{cm}$ horizontal by $12\,\mathrm{cm}$ vertical.  The guide has a curved section shortly after the moderator that prevents direct line of sight from the experiment to the moderator, reducing the fast neutron and gamma backgrounds in the experiment.

Of the four beam chopper stations indicated in Fig.~\ref{fig:FnPBSchem}, the first two stations, after the moderator, at $5.5\,\mathrm{m}$ and $7.5\,\mathrm{m}$, have been equipped with time-of-flight frame definition choppers. The main function of the choppers is to select the neutron energy range in each pulse for the experiment with a goal to optimize the statistics and the capture range in the chamber. The PV effect is independent of the neutron energy. Each chopper consists of a large carbon fiber wheel coated with $^{10}$B, which has a high neutron absorption cross-section for low energy neutrons, and each wheel  has a single opening for neutrons to pass the disk without being absorbed.
Each chopper is synchronized to the arrival of proton pulses on the Mercury target, allowing for an accurate selection of the TOF frame in each neutron pulse. The second function of the choppers was, to prevent the neutrons in a given pulse to mix with neutrons of the previous pulse, which is possible due to the $20$~m distance of the n$^3$He apparatus from the moderator. Figure Fig.~\ref{fig:TwoPulsePreampOut} shows the effect of the choppers on the TOF spectrum and pulse separation. For n$^3$He, the chopper phasing was adjusted such that neutrons with energies between about $2~\mathrm{\text{meV}}$ and $9~\mathrm{\text{meV}}$ were selected, as indicated by the grayed area in Fig~\ref{fig:SNSFullPulse}. The corresponding neutron fluence, after integrating over the selected neutron energy range and the beam profile, was about $1.8\times 10^{10}~\mathrm{n/s/MW}$~\cite{Tang}. The average delivered proton beam power varied over the duration of the experiment, from $0.7$ to $1.4~\mathrm{MW}$.

\subsection{The n$^3$He apparatus} \label{sect:n3Heoverview}
\begin{figure*}[t]
  \begin{center}
  \includegraphics[width=0.9\textwidth]{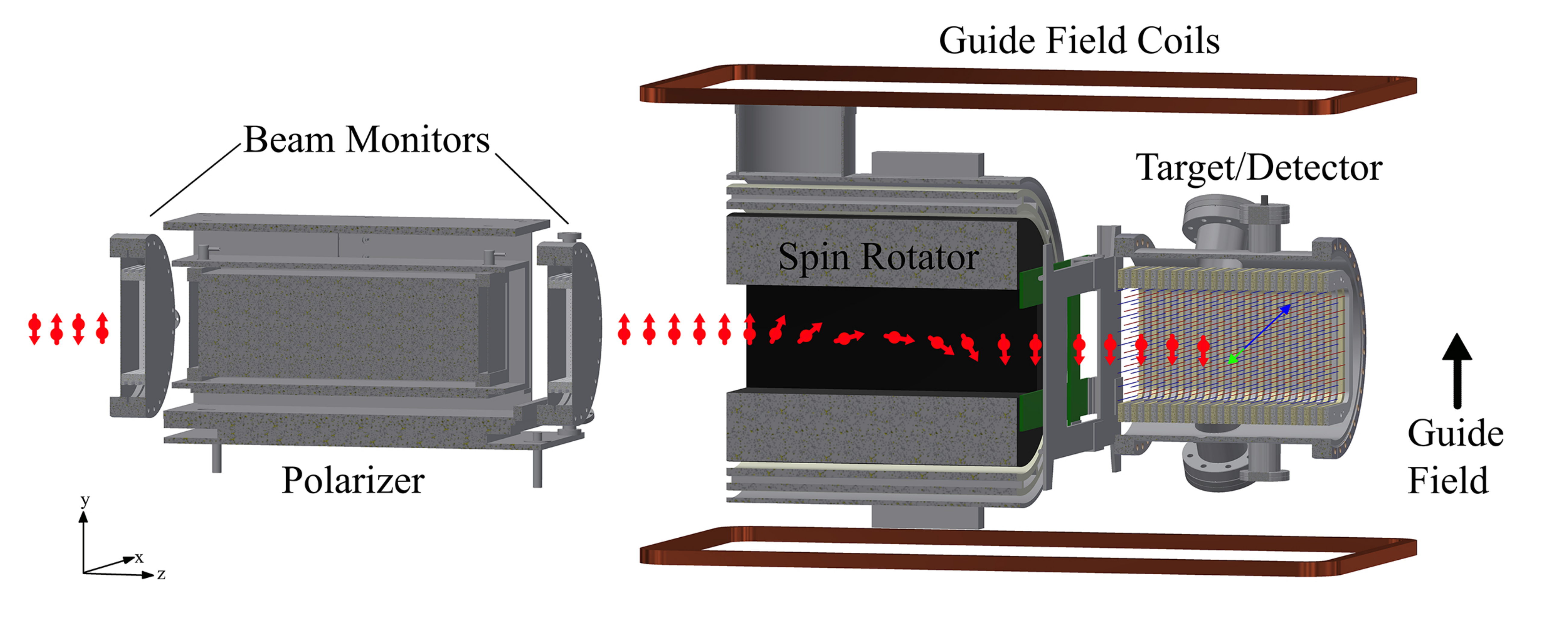}
    \caption{Illustration of the n3He apparatus. Neutrons enter from the left and travel in the $+\hat{z}$ direction. The beam monitor measures the relative neutron beam intensity and pulse shape. Neutrons exit the supermirror with spins aligned upward (along $+\hat{y}$). The RF spin rotator reverses the spin direction every other beam pulse.  Before entering the target the beam was collimated. Neutrons are captured in the target-detector chamber by $^{3}\mathrm{He}$, producing a proton and a triton per capture (the blue and green arrows respectively).}
    \label{fig:n3HeSetup}
  \end{center}
\end{figure*}

The experimental setup of the n$^3$He experiment is shown schematically in Fig.~\ref{fig:n3HeSetup}. Starting with the exit of the neutron guide, in beam direction $+\hat{z}$), the experiment consisted of a beam monitor, a supermirror polarizer, a resonant RF spin rotator, a jaw collimator system, a holding field and an ion chamber that functioned as an active target. A set of four race-track shape magnetic field coils were used to produce a 10 Gauss homogeneous field, to hold the neutron polarization after the polarizer. The holding field direction was carefully aligned to the $+\hat{y}$ direction (the neutron polarization axis). The beam monitor was a $^3$He ionization chamber with low neutron capture efficiency, interrupting only a few \% of the beam. A supermirror polarizer (SMP) \cite{SwissNeutronics} was used to polarize the neutrons by spin dependent reflection~\cite{Mezei,Balascuta2012137}. A detailed description of the experiment as well as some preliminary analysis can be found in~\cite{thesis:MMcCrea,thesis:LKabir,thesis:Coppola,Thesis:EricPlemmons}.

%\begin{figure}[htb]
%  \begin{center}
%  \includegraphics[width=0.9\columnwidth]{figures/n3HeExpLayout.eps}
%    \caption{Schematic of the n$^3$He apparatus. The neutrons enter from the left and travel in the $+\hat{z}$ direction. The $^{3}\mathrm{He}$ beam %monitor measured relative neutron beam intensity and pulse shape. The initial neutron polarization after the supermirror polarizer is vertically up %(along $+\hat{y}$) and is reversed for every other beam pulse by the resonant RF spin rotator. Before entering the target the beam was collimated. %The neutrons are captured in the target-detector chamber by $^{3}\mathrm{He}$, producing one proton and one triton per capture.}
%    \label{fig:n3HeSetup}
%  \end{center}
%\end{figure}

The initial neutron polarization after the SMP was oriented vertically up ($+\hat{y}$ direction in Fig.~\ref{fig:n3HeSetup}). The polarization was reversed every other pulse by the resonant spin rotator (RFSR) - with the RF field on. The amplitude of the RF field was carefully matched to the energy of the incident neutrons to effectively reverse the polarization of each neutron in a pulse~\cite{Seo2008}. For production data taking, the neutron beam was collimated in front of the target-detector chamber, to $8~\mathrm{cm}$ in the $\hat{y}$-direction and $10~\mathrm{cm}$ in the $\hat{x}$-direction. The neutrons were captured in the target-detector by $^3\mathrm{He}$ gas and the protons and tritons from the capture reactions produced the ionization signal. The details of the chamber are discussed in sec.~\ref{sec:TargetDesign}.

\subsection{Measurement Principle}

As mentioned in the introduction, the measurement is based on the $^3\mathrm{He}(\mathrm{n},\mathrm{p})^3\mathrm{H}$ reaction. For cold neutrons the capture reaction on $^3\mathrm{He}$ occurs essentially 100\% of the time through the $^3\mathrm{He}(\mathrm{n},\mathrm{p})^3\mathrm{H}$ channel, with a very small branching ratio for radiative capture $^3\mathrm{He}(\mathrm{n},\gamma)^{4}\mathrm{He}$. The corresponding unpolarized capture cross-sections are $\sigma_c = \sigma_o \frac{\mathrm{v_o}}{\mathrm{v}}$ ($\sigma_o = 5327\pm10~\mathrm{b}$) and $\sigma_{rc} = 56\pm6~\rm{\mu b}$~\cite{WOLFS}, where $\sigma_o$ and $\mathrm{v_o} = 2200~\mathrm{m/s}$ are the capture cross section and the speed for thermal neutrons. For the cold neutron energies used in this experiment, the $^{4}\mathrm{He}^{*}$ decays essentially at rest, with negligible  recoil effects, and the unpolarized cross-section is spherically symmetric. As a result of the very large capture cross-section, the vast majority of the neutrons are absorbed by the $^3\mathrm{He}$ in the chamber, even for relatively small volumes at pressures close to atmosphere (see Fig.~\ref{fig:nCapDist}). This allowed for a very precise (high statistics) determination of the experimental observable, the proton asymmetry, over a relatively modest running time.

In this experiment, the asymmetry in the emission direction of the proton ($\hat{k}_p$), with respect to the neutron spin ($\hat{s}_n$) was measured. The corresponding differential cross section is given by
\begin{equation}\label{eq:DiffCrossSection}
\frac{d\sigma}{d\Omega} = \left(\frac{d\sigma}{d\Omega}\right)_\mathrm{c}\left(1+A_{_{\mathrm{PV}}}\cos\theta_y + A_{_{\mathrm{PC}}}\cos\theta_x \right)~.
\end{equation}
Here, $\left(\frac{d\sigma}{d\Omega}\right)_\mathrm{c}$ is the unpolarized neutron capture cross-section, $A_{\mathrm{PV}}$ is the parity-violating (PV) asymmetry and $A_{\mathrm{PC}}$ is the parity-conserving (PC) asymmetry. The PV asymmetry is a result of the correlation $\hat{s}_n\cdot \hat{k}_p = \cos\theta_y$, while the PC asymmetry is a result of the correlation $\left(\hat{s}_n\times\hat{k}_n\right)\cdot \hat{k}_p = \cos\theta_x$ For the definition of the PC correlation, we are generally using the coordinate system of Ohlsen and Keaton~\cite{Ohlsen}, but with the azimuthal angle $\phi$ measured from the spin axis $y$ to the scattering normal, $\vec{n} = \hat{k}_n \times \hat{k}_p$. In standard spherical coordinates $\cos\theta_y = \sin\theta\sin\phi$ and $\cos\theta_x = \sin\theta\cos\phi$.

Referring to Fig.~\ref{fig:n3HeSetup}, since the beam polarization is purely transverse, along $\pm \hat{y}$ (with the beam momentum equal to $k_n\hat{z}$), the vector $\hat{s}_n\times\hat{k}_n$ lies along the $\pm\hat{x}$ direction and the angles $\theta_y$ and $\theta_x$ give the proton momentum direction with respect to the $\hat{y}$ and $\hat{x}$ axes respectively. Therefore, when the neutron spin is reversed, using the spin rotator, the sign of the correlation terms flips along the corresponding axis. The PV asymmetry can be extracted by measuring the signal with separate detectors in the upper and lower hemispheres  ($\pm\hat{y}$), rejecting the PC asymmetry by using continuous detectors extending symmetrically in the horizontal($\pm\hat{x}$) direction, around the beam centroid. We call this the up-down (UD) measurement mode. Likewise, the PC asymmetry can be measured by implementing separate detectors in the left and right hemispheres, while rejecting the PV asymmetry with continuous symmetric detectors in the vertical ($\pm\hat{y}$) direction, which we refer to as the left-right (LR) mode. As mentioned above, the neutron beam polarization was reversed for every other pulse ($30$~Hz) and asymmetries were formed for each pair of opposite polarization-states (indicated by $\pm$ superscript):
\begin{equation}\label{eq:Asy1}
  \epsilon P~A_{_{\mathrm{PV/PC}}} = \frac{\sigma^{+} - \sigma^{-}}{\sigma^{+} + \sigma^{-}} ~.
\end{equation}
Here, the factors $\epsilon$ and $P$ represent the polarization reversal efficiency and beam polarization, respectively.

As explained in detail in the next section, the target-detector chamber was primarily designed to make the PV asymmetry measurement (UD mode), while rejecting the PC asymmetry. However, using construction and alignment symmetry, a simple $90^\circ$ rotation of the chamber about the beam axis allowed the switch to a configuration in which the PC asymmetry could be measured (LR mode), while rejecting the PV asymmetry. The separate measurement of the PC asymmetry was used to check for systematic effects and to verify operation of the chamber, since the PC asymmetry was expected to be non-zero and large.

\section{The Target-Detector Chamber}\label{sec:TargetDesign}
\begin{figure}[ht]
  \centering
  \includegraphics[width=0.75\columnwidth]{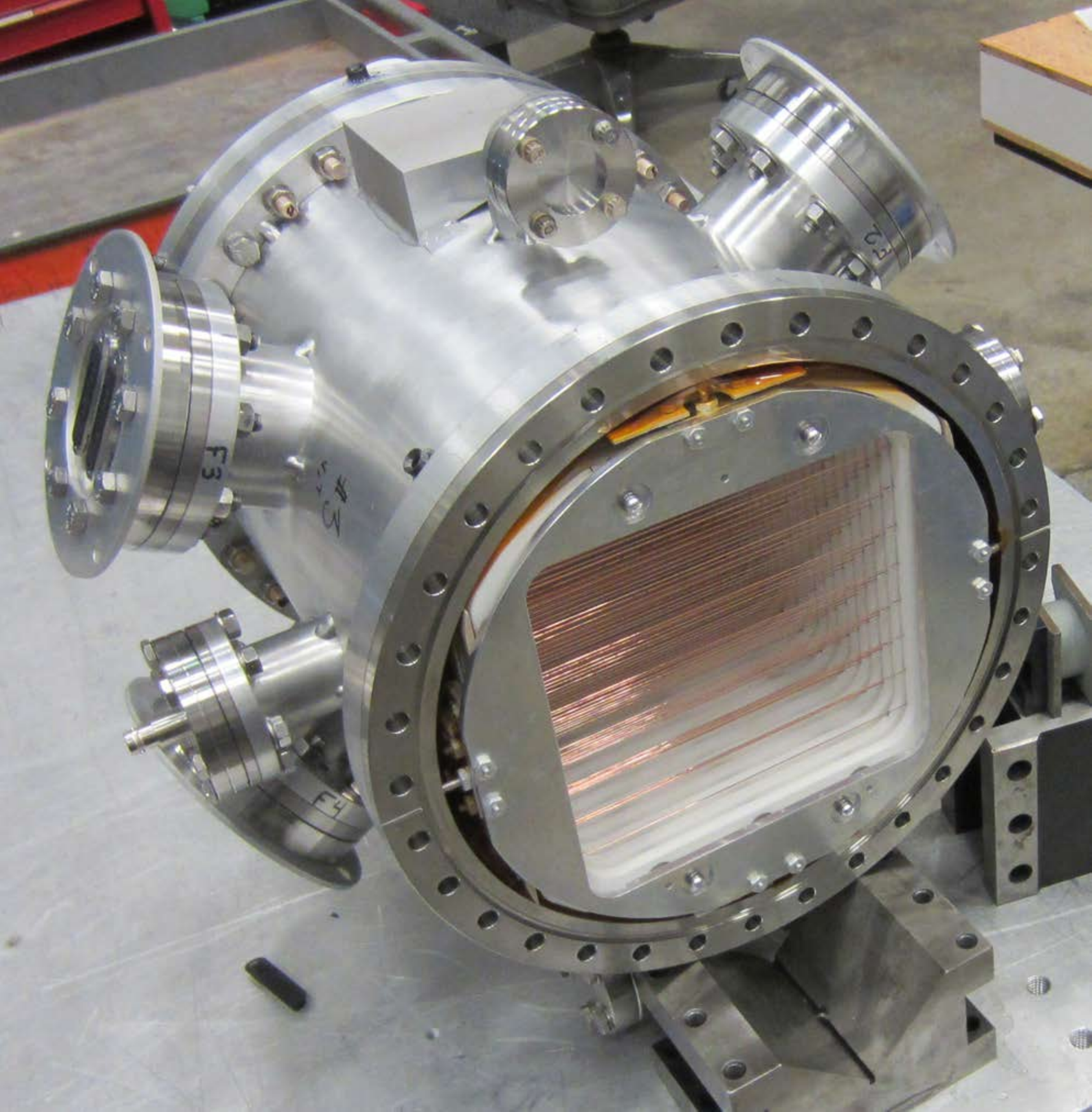}
  \caption{Assembled target-detector chamber. The chamber consisted of an all aluminum cylindrical housing (except for the conflat flange knife edges, which were friction welded onto the aluminum), several signal, high voltage and gas feedthroughs, and the wire frame stack. The chamber windows (not shown here) consisted of $0.5~\mathrm{mm}$ thick aluminum machined down from $12~\mathrm{inch}$ aluminum blank conflat flanges. The chamber had an inner diameter of $254~\mathrm{mm}$ and a length of $338~\mathrm{mm}$.}
  \label{fig:chamberfig}
\end{figure}
\begin{figure}
  \centering
  \includegraphics[width=0.4\textwidth, keepaspectratio]{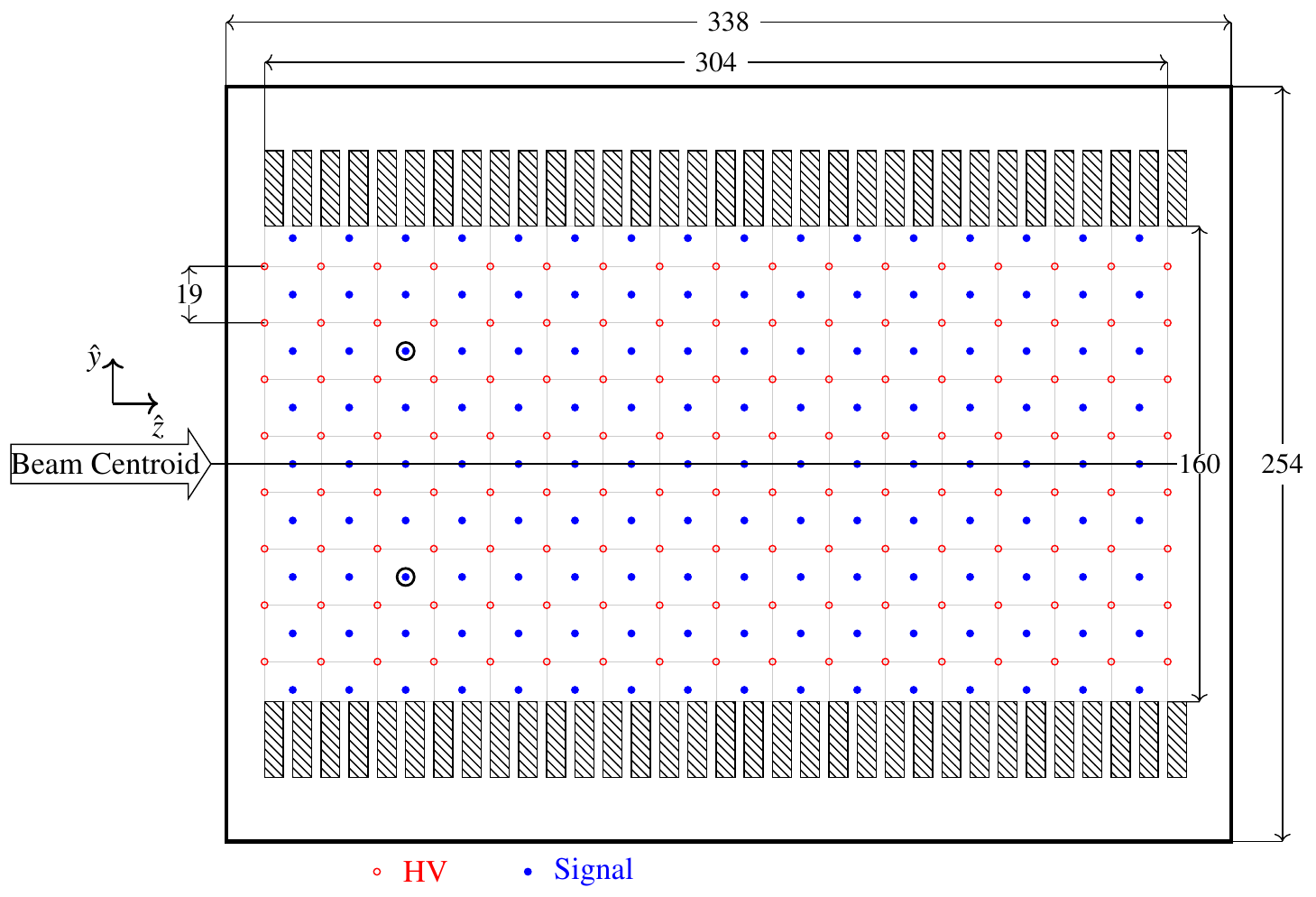}
  \caption{Schematic of the wire layout in the n$^3$He chamber, shown in cross section through the chamber’s central axis. Each colored point represents one wire, running perpendicular to the page, red indicates HV wires and blue charge collection wires. During operation the wires were aligned parallel to either the $\hat{x}$ (horizontal) or $\hat{y}$ (vertical) beam-line axis to measure either PV or PC asymmetries respectively. Linear dimensions are in millimeters. The circles around two of the signal wires indicate an example of a wire pair, which are wires that mirror each other through the $(x,z)$ plane centered on the beam centroid axis. There are 64 such pairs, which are used to compute the measured total asymmetry (see text).}
    \label{fig:wirelayout}
\end{figure}

The target-detector was implemented as a multi-wire chamber using a pure $^3$He fill gas. The primary requirements for the chamber were a high absorption of neutrons, a long range-out path of the protons, efficient, low noise signal collection, low background production from neutron capture and outgassing of the chamber materials, and a high degree of symmetry and alignment. A photo of the assembled target-detector chamber (without beam-windows) is shown in Fig.~\ref{fig:chamberfig}. It consisted of a cylindrical housing with thin windows, almost entirely made from aluminum (except for the conflat flange knife edge), with various signal, high voltage, and gas feedthroughs. The housing cylinder had a length of $338~\mathrm{mm}$ and an inner diameter of $254~\mathrm{mm}$. During operation, the entire chamber was filled with $^3\mathrm{He}$, at a pressure of $0.47~\mathrm{atm}$. The active detection volume of the chamber was defined by a stack of 31 alternating high voltage and signal wire planes. The wire layout is illustrated in Fig.~\ref{fig:wirelayout}. The inner (active) volume of the frame stack was $304~\mathrm{mm}$ long, $140~\mathrm{mm}$ wide (beam left-right), and $160~\mathrm{mm}$ tall (beam up-down direction). The width and height of the frame stack were designed to cover the beam profile at the detector location, taking into account beam divergence and collimation. For the PV asymmetry measurements, the chamber was oriented such that the wires were horizontal, as shown in~Fig.~\ref{fig:chamberfig}.

\subsection{Signal Generation, Chamber Segmentation, and Asymmetry Sensitivity}

The $Q$-value of the neutron-$^{3}$He capture is  $765\,\mathrm{keV}$ which converts to kinetic energy shared between the final state proton ($571\,\mathrm{keV}$) and triton ($194\,\mathrm{keV}$). The chamber signal was formed by the ions, produced by the proton and triton, as they travel through the $^3$He gas.  Using a pure $^3$He fill gas ensures that the chamber can act as the target and detector at the same time.  While the use of a gas mixture (e.g. with $\mathrm{N_2}$) could have improved the high voltage and signal characteristics of the target-detector chamber, additional gases would have produced an irreducible background and systematic effect from neutron capture on the corresponding elements, significantly complicating the $^3$He asymmetry extraction.

\begin{figure}[ht]
  \centering
    \includegraphics[width=0.9\columnwidth]{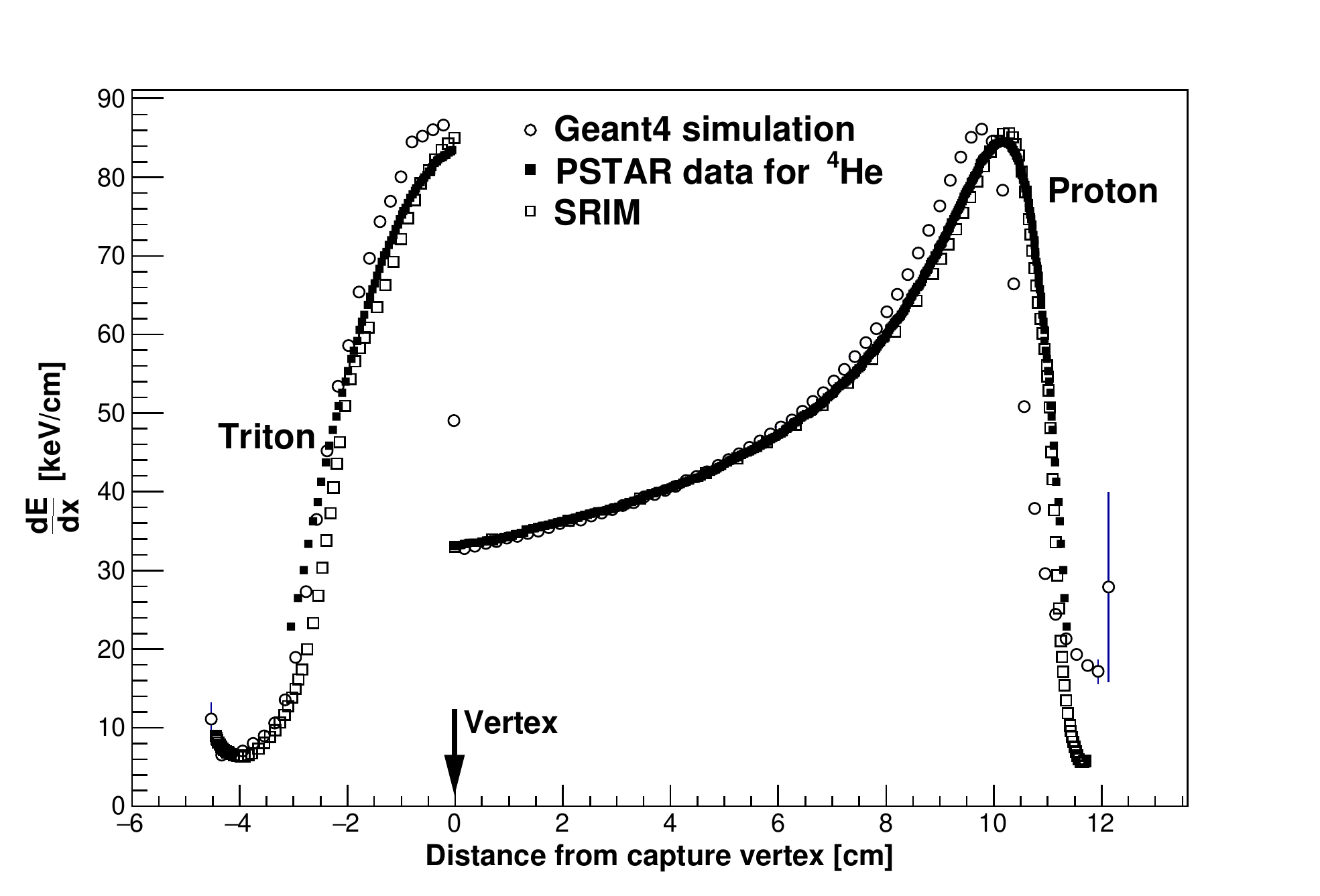}
  %created with root script:
  %/Users/mark/Project_Information/n3He/Design Calculations/StopPow/ReadCSV.C
  \caption{Energy deposited by the proton and triton per $0.1\,\mathrm{cm}$ step in $0.47\,\mathrm{atm}$ He gas. The Geant4 and SRIM simulations were done with $^{3}\mathrm{He}$ as a fill gas. The PSTAR data \cite{PSTAR} was available only for $^{4}\mathrm{He}$.}
  \label{fig:n3HeBraggCurve}
\end{figure}

Due to the high neutron capture rate in the target, the chamber had to be operated in signal integration mode, rather than pulse counting mode, as further explained below. This meant that both, the proton and the triton signal where simultaneously measured and could not be discriminated. Since the proton and triton are emitted back-to-back, a consequence of the integration was a reduction of the measured asymmetry. To nonetheless be able to make the proton asymmetry measurement, the most important property of the interactions in the chamber was that the proton range be significantly larger than the mean free path of the neutrons ($\sim 5~\mathrm{cm}$) or the range of the triton ($\sim 3$ cm).
\begin{figure}[h]
  \begin{center}
  \includegraphics[width=0.9\columnwidth]{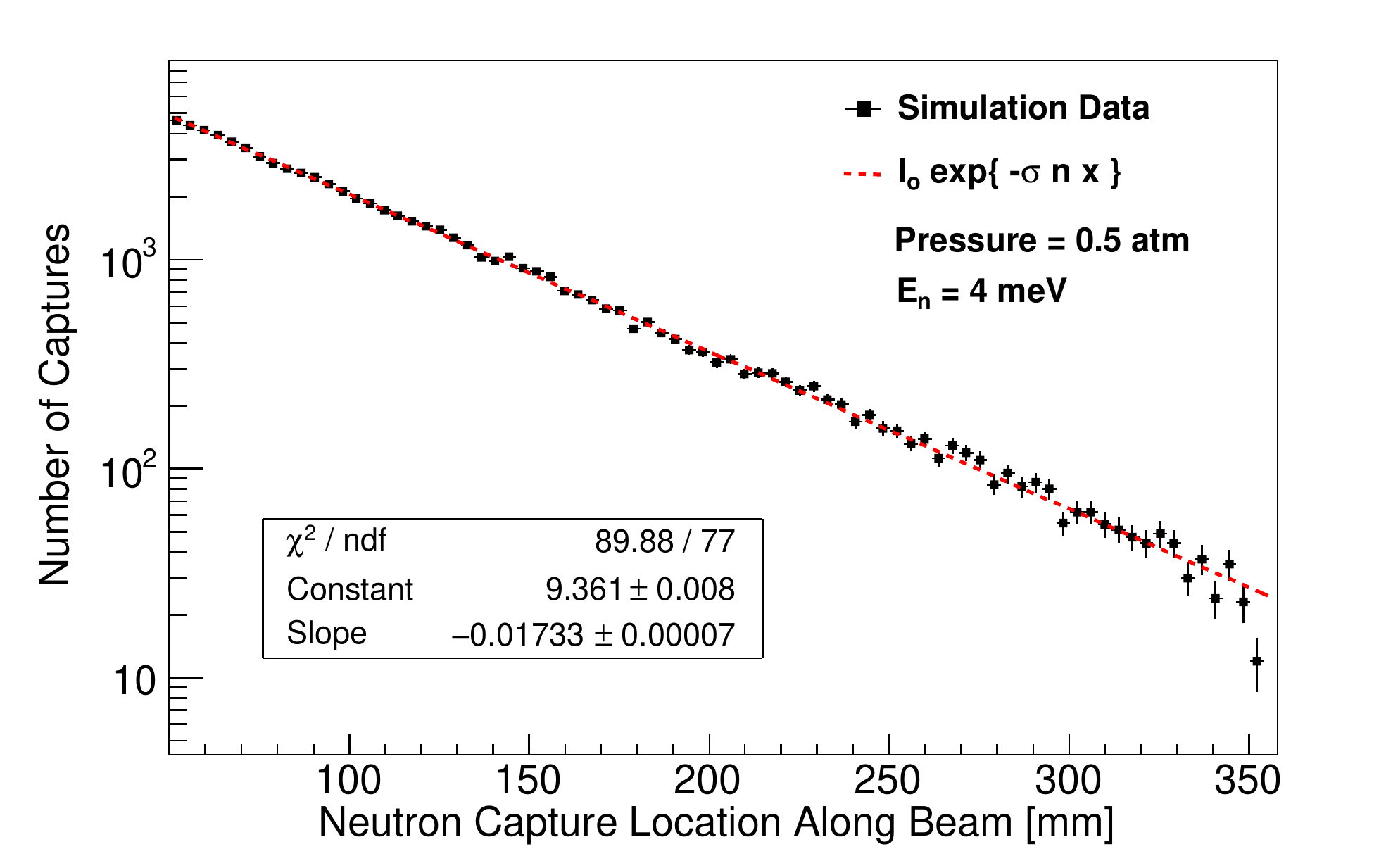}
    \caption{Simulated capture distribution for $4~\mathrm{meV}$ neutrons in the target-detector chamber, along the beam direction. The capture cross-section is extracted from the slope of the fit to the simulation data and compared to the known cross-section, as provided by the ENDF libraries. This serves as the validation of the Geant4 simulation, for cold neutron capture.}
    \label{fig:nCapDist}
  \end{center}
\end{figure}
Fig.~\ref{fig:nCapDist} shows that the vast majority of the neutrons are captured in the front of the chamber, as expected, and Fig.~\ref{fig:n3HeBraggCurve} shows the ionization energy deposited by the proton and triton, as a function of track length. Since the proton energy is large, its energy deposit in He gas has the typical Bragg features, as shown in the figure. The triton energy, on the other hand, is so small that the chamber sees only the tail end of the Bragg curve. Due to this, the length of the proton path, and the charge collection volumes defined by the signal wire planes, the chamber was mostly sensitive to the proton emission direction. The triton signal leads to a known (calculable) dilution of the proton asymmetry, which is taken into account in the data analysis, through the use of geometry factors, as explained further on in the paper.

The efficiency of the wire chamber depends in a complicated way on the proton emission angles, neutron energy, and wire plane spacing. Detailed Monte-Carlo (MC) simulations were used to determine the optimal neutron wavelength interval, the optimum number and separation of wire planes, and the optimum gas pressure, as well as the suitable wire bias voltage. Some of these simulations are briefly described below. The MC was used to predict the signal RMS width above neutron counting statistics $\sigma_{_{\mathrm{RMS}}}/\sqrt{N}$ and the corresponding needed running time of the experiment. Based on this, the target-detector active volume was separated into 144 detection regions, using evenly spaced signal wires surrounded by high voltage wires, as shown in Figs.~\ref{fig:wirelayout} and~\ref{fig:chamberfig}.
\begin{figure}[]
  \begin{center}
  \includegraphics[width=0.9\columnwidth]{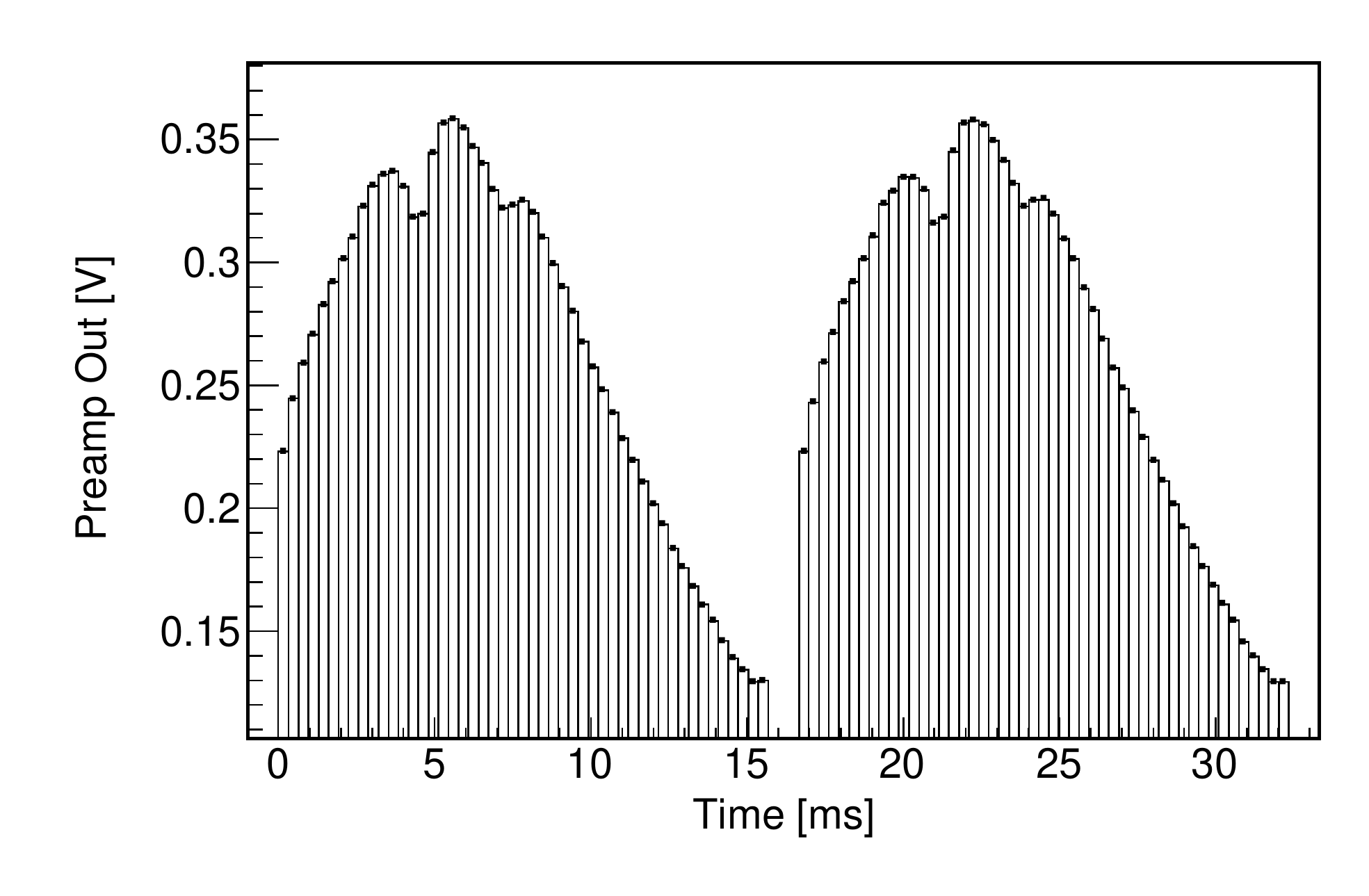}
    \caption{Preamplifier output voltage for one sense wire in the center of the chamber for two consecutive pulses. The effect of the beam choppers on the spectrum is evident in the gap between successive pulses. The signal for each pulse was separated into 49 time bins of about 0.32 ms width. The gap between pulses was about 0.98 ms long.}
    \label{fig:TwoPulsePreampOut}
  \end{center}
\end{figure}
Each of these regions, which we subsequently refer to as cells, define a volume within which the ionization charge is collected. As discussed below, simulations have shown, that the charge produced in a given cell was essentially collected completely within that cell. To measure the PV asymmetry, the target-detector chamber had to have finite resolution in the direction parallel to the magnetic field that defined the neutron beam polarization in the capture. To measure the PC asymmetry, the target-detector chamber had to have finite resolution in the direction perpendicular to the magnetic field. Therefore, for the PV asymmetry measurement, the wires ran the full width of the frame (which was beam left-right with the wires oriented horizontally) or, for the PC asymmetry measurement, the wires ran beam up-down (oriented vertically). Thus, by reading out each signal wire individually, the position of ionization in the target-detector chamber could be measured in two dimensions (along the beam direction and in either one of the transverse directions).

\subsection{Integrating Signal Readout and Asymmetry Formation}

Equation~\ref{eq:DiffCrossSection} provides the starting point for the calculation of the wire yields. Taking into account the fact that the target-detector, the wire cells, and the beam have a finite size, and the fact that the beam has a finite polarization, the signal wire voltage yield can then be written as
\begin{equation}\label{eq:wireyield}
Y^{\pm}_i = R_i I_{\mathrm{o}}\left(1\pm\epsilon P \left(A_{_{\mathrm{PV}}}G^{PV}_{i} + A_{_{\mathrm{PC}}}G^{PC}_{i}\right)\right)+p_i~.
\end{equation}
%\begin{equation}
%Y^{\pm}_{\mathrm{k}} = I^{\pm}_{\mathrm{o}} R_{\mathrm{k}}(1\pm \epsilon P G_{\mathrm{k,PV}} A_{\mathrm{PV}} \pm \epsilon P G_{\mathrm{k,PC}} %A_{\mathrm{PC}}) + p^{\pm}_{\mathrm{k}}\label{eq:wireyield}~.
%\end{equation}
Here, $I_{\mathrm{o}}$ is the neutron beam intensity in units of $[\mathrm{n/s}/\text{meV}]$ (integrated over the beam size at the target-detector), $R_i$ incorporates the wire ``\textit{gain}'' for wire cell $i$, in units of [C/n] and the gain of the amplifier in units of [Ohm], $\epsilon$ is the unit-less fractional efficiency of the spin flipper, $P$ is the absolute value of the fractional beam polarization, and $p_i$ the electronic pedestal. The $G^{PV/PC}_i$ are so-called geometry factors, which replace the $\cos{\theta_y}$ and $\cos{\theta_x}$ factors in eqn.~\ref{eq:DiffCrossSection}, for a finite size target-detector cell, finite beam geometry, and a neutron capture distribution that varies with position inside the target-detector and with neutron energy (see sec.~\ref{sec:Sims}).

\begin{figure}[htb]
  \begin{center}

    \includegraphics[height=0.74\columnwidth, angle=90]{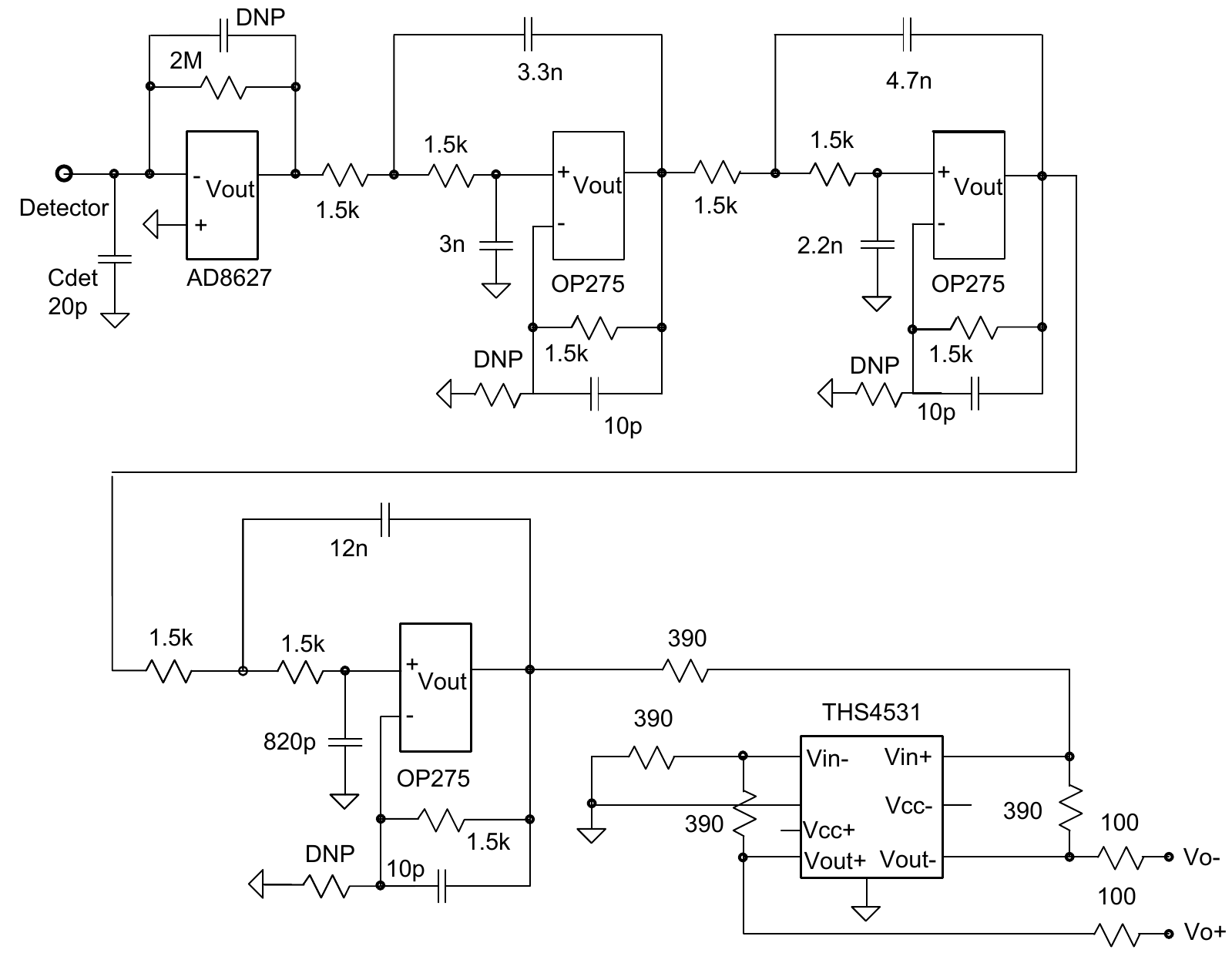}
    \caption{A schematic of one channel of the amplifier electronics.}
    \label{fig:circuit}
  \end{center}
\end{figure}

The current from each signal wire was read and amplified by a fast, 500~MHz, AD8627 JFET current-to-voltage amplifier that has a low intrinsic current noise density of 0.4 fA/$\sqrt{\rm{Hz}}$. The overall gain of the amplifier circuit is set by the 2~M$\Omega$ feedback resistor as shown in Fig.~\ref{fig:circuit}.
After amplification the signal is shaped by an active RC-filter and the low-noise THS4130 amplifier provides a high-speed differential signal driver to the ADC input. Three signal shaping stages each with an $RC$ time constant of $15\;\mathrm{ns}$ occurred between the two amplifiers.
The circuit was analysed with LTspice, and showed a very flat frequency response up to the $20~\mathrm{kHz}$ $f_{3dB}$ cutoff frequency that was selected to be well below the Nyquist limit of the $50~\mathrm{kSps}$ ADC sampling rate. The Johnson-Nyquist thermal voltage noise density in the output of the amplifier chain was $180~\mathrm{nV/\sqrt{Hz}}$ corresponding to the $90~\mathrm{fA/\sqrt{Hz}}$ that is produced by the $2~\mathrm{M\Omega}$ feedback resistor. The output current density is significantly larger than the intrinsic noise of $0.4~\mathrm{fA/\sqrt{Hz}}$ of the AD8627 amplifier, but more importantly, the $90~\mathrm{fA/\sqrt{Hz}}$ is significantly smaller than the average shot noise density of the experiment which allowed in situ measurement of any false asymmetry of instrumental origin on a short time scale, compared to the running time of the experiment, as discussed in section \ref{sec:InstrumentalAsym}. Each amplifier output was sampled at $50~\mathrm{kSps}$, by a 24-bit ACQ435ELF-24 (Texas Instrument ADS1278 delta-sigma~\footnote{http://www.ti.com/lit/ds/symlink/ads1278.pdf}) ADC, mounted on an ACQ1002R-2 carrier board made by D-tAcq Solutions~\footnote{http://www.d-tacq.com/}. Further details about the data acquisition system can be found in~\cite{thesis:LKabir}.

Sixteen consecutive samples were averaged to form a $0.32~\mathrm{ms}$ long time bin and each chopped neutron pulse consisted of $49$ of these 0.32~ms long time bins, as shown in Fig.~\ref{fig:TwoPulsePreampOut}.  For the asymmetry analysis, the signal was integrated in each pulse, from time bin 5 to 44. The measured asymmetries were constructed from consecutive neutron beam pulses with opposite spin directions for each sense wire in the chamber, as described below. The 144 amplifier channels were distributed between four 36-channel motherboards that were mounted into covered Al boxes that were mounted directly onto four vacuum ports of the detector chamber to minimize the length of the transmission lines, to provide RF shielding, and to provide a cooling confinement for the electronics by flowing $\sim10~\mathrm{liter/min}$ of dry nitrogen through the box.

Asymmetries are calculated from pairs of conjugate wires (see Fig.~\ref{fig:wirelayout}) which have geometry factors that are equal in magnitude and opposite in sign. Wire pairs are used for the asymmetry calculation, to remove false asymmetries from beam fluctuations, which can be of $O(10^{-5})$.  The corresponding measured (PV) asymmetry from a given conjugate wire pair $(\mathrm{k,k}^*)$, in a single pair of pulses with opposite neutron spin $(\pm)$, is given by
\begin{eqnarray}\label{eqn:Asym} \nonumber
A^{meas}_{\mathrm{k}} &=& \frac{Y_{\mathrm{k}}^{+} - Y_{\mathrm{k}}^{-}}{Y_{\mathrm{k}}^{+} + Y_{\mathrm{k}}^{-}}+\frac{Y_{\mathrm{k}^*}^{+} - Y_{\mathrm{k}^*}^{-}}{Y_{\mathrm{k}^*}^{+} + Y_{\mathrm{k}^*}^{-}} \\
    &=&  2\epsilon P G^{PV}_{\mathrm{k}} A_{\mathrm{PV}} + A^{ped}_{\mathrm{k}}-A^{ped}_{\mathrm{k^*}}~.
\end{eqnarray}
Here $A_{\mathrm{ped}}$ is the asymmetry in the pedestal for a given wire. For a given pulse pair, the pedestal asymmetries are generally non-zero at the $O(10^{-3})$ level, but with good noise properties and proper decoupling of the electronics from the spin state signals, they will average to zero, over many measured pulse pairs. This was verified during data taking, by performing many periodic pedestal runs
(see sec.~\ref{sec:InstrumentalAsym}).

\section{Chamber Design and Construction Details}\label{sec::DesConst}

The target-detector chamber had to be designed to reduce backgrounds and instrumental systematic effects as much as possible, while maximizing the signal yield and asymmetry sensitivity. This placed specific requirements on materials and design details. This section describes in detail, the design and construction of the chamber. A comprehensive account of the chamber design, construction, and commissioning can be found in~\cite{thesis:MMcCrea}.

\subsection{Target-Detector Housing}\label{sec:TargetHousing}

As mentioned in sec.~\ref{sec:TargetDesign}, with the exception of the conflat knife edge for the target windows and feedthroughs, the chamber is entirely made from aluminum. This material choice is primarily driven by the requirement to have non-magnetic materials to prevent distortion of the holding field that is used to maintain the neutron polarization after the spin-flipper.  Aluminum also has a low neutron capture cross section of $0.231$ barn for thermal neutrons ($v_n = 2200\,\mathrm{m/s}$) and a low scattering cross section of $1.503$ barn. In addition, $^{28}\mathrm{Al}$ (from neutron capture on $^{27}\mathrm{Al}$) has a short half-life of $2.245\,\mathrm{min}$, which means that the background from activation on the chamber walls quickly reaches a steady state. $^{28}\mathrm{Al}$ beta decays to $^{28}$Si, followed by several MeV-level gamma rays.  The gamma rays will not create significant ionization in the target chamber, the only channels being pair production and Compton scattering.  The beta decay electrons have sufficient energy to exit the aluminum window and ionize the fill gas, but due to the high Q-value ($4.642$~MeV) and small mass of the electron, the $dE/dx$ for betas is about 100 times smaller than the corresponding values for the proton and triton of the ${^3}He(n,p)^3H$ reaction. Finally, aluminum is also a low out-gassing material. The target chamber housing was made by Atlas Technologies~\footnote{http://www.atlasuhv.com/}.

%The target chamber body is an aluminum cylinder with a $25.4\,\mathrm{cm}$ interior diameter and $0.635\,%\mathrm{cm}$ thick walls.  The end flanges were 12 inch conflat flanges.  The end windows for the target chamber were $1\,\mathrm{mm}$ aluminum windows to maximize neutron transport into the target chamber.  Fig.~\ref{fig:TargetHousing} shows the target housing before assembly.
%\begin{figure}[tbp]
%  \centering
%    \includegraphics[width=0.45\textwidth]{figures/IonChamber-Assembly-TargetHousing}
%  \caption{The aluminum target housing and end windows for the the n3He target.}
%  \label{fig:TargetHousing}
%\end{figure}

Two pairs of $4.5$ inch conflat nipples on opposite sides of the chamber were used with Lesker $2\times25$ pin Dsub vacuum feedthroughs, to read out the signal wires.  Two $2.75$ inch feedthroughs on opposite sides of the target chamber were used for the HV supply feedthroughs.  One of the two $90^\circ$ feedthroughs was used for the gas fill. Fig.~\ref{fig:chamberfig} shows the target housing with one end window removed. All of the conflat flanges on the target housing and end windows are bimetal flanges~\cite{AtlasBimetalFlange}.  These flanges are created by explosion welding a stainless steel sheet to an aluminum sheet from which the flange is then machined.  The knife edges are formed out of stainless steel to make them more resistant to damage while minimizing the amount of steel in the target.  Stainless steel is also a low outgassing material.

\subsection{Frame Stack Design}\label{sec:FrameStackDesign}

\begin{figure}[tbp]
  \centering
  \includegraphics[width=0.48\textwidth, height = 0.3\textheight, keepaspectratio]{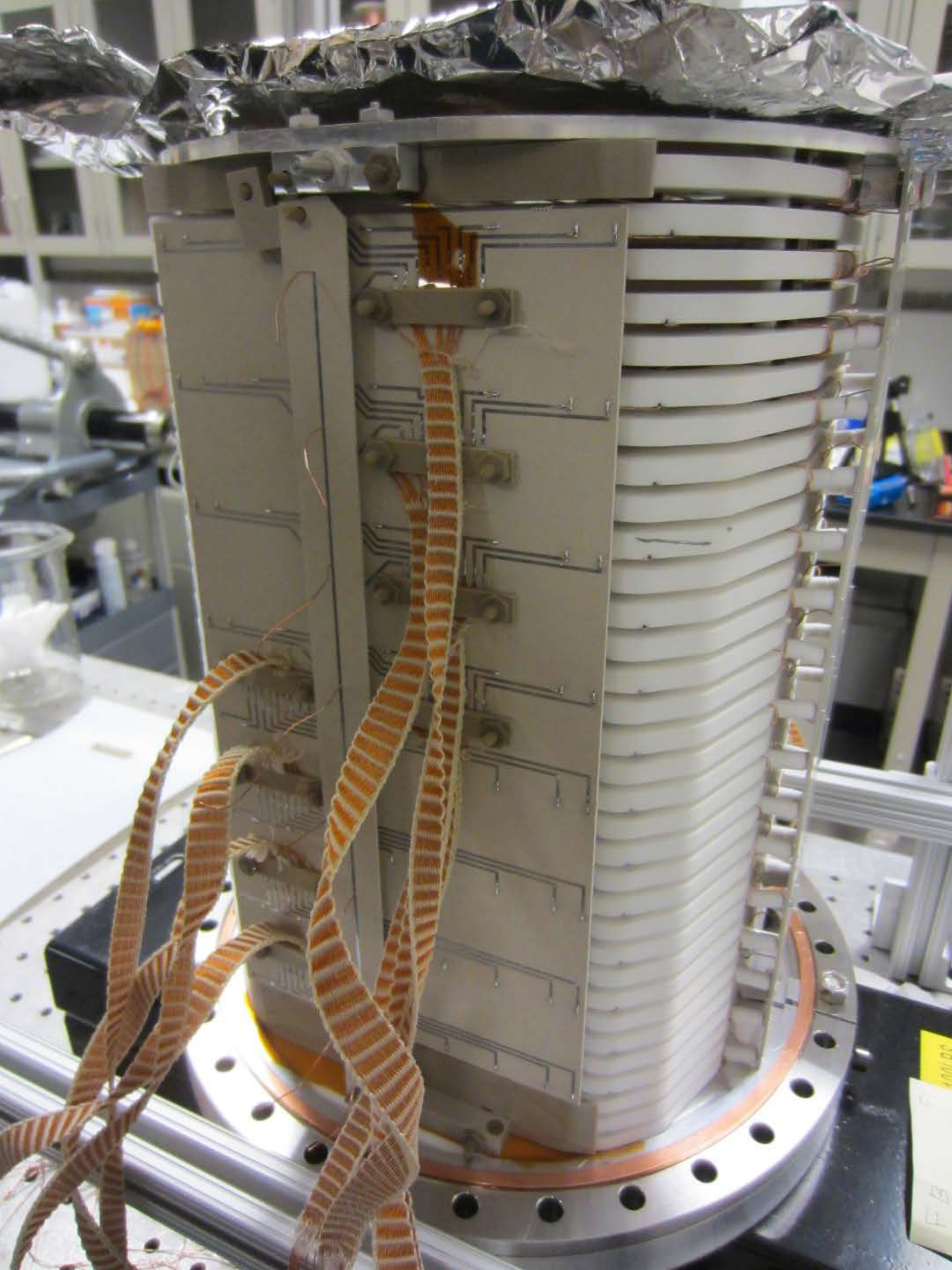}
  \caption{Assembled target frame stack. The traces for the wire signals can be seen, connected to the kapton ribbon cables that were later connected to the 25 pin D-Sub plug feedthroughs. The frame material was macor and the trace boards were made from PEEK.}
  \label{fig:FrameStack}
\end{figure}

As explained earlier, the active (signal detection) volume of the chamber was separated into 144 volumes (cells), using the configuration shown in fig.~\ref{fig:wirelayout}. Each wire plane (defined by the wires in a plane with a normal that is parallel to the beam direction) was mounted on separate wire frame. The entire frame stack then consisted of alternating high voltage (HV) and signal wire plane frames. The assembled stack is shown in fig.~\ref{fig:FrameStack}. This design is based on various simulations and calculations, optimized with respect to the ionization response in each wire plane, the sensitivity to the asymmetry in each wire plane, the statistical error, the correlations between the wire cells, and the ability to measure the asymmetry while canceling out the false asymmetry from beam fluctuations.

\begin{figure}[tbp]
  \centering
    \includegraphics[width=0.48\textwidth, height=0.3\textheight, keepaspectratio]{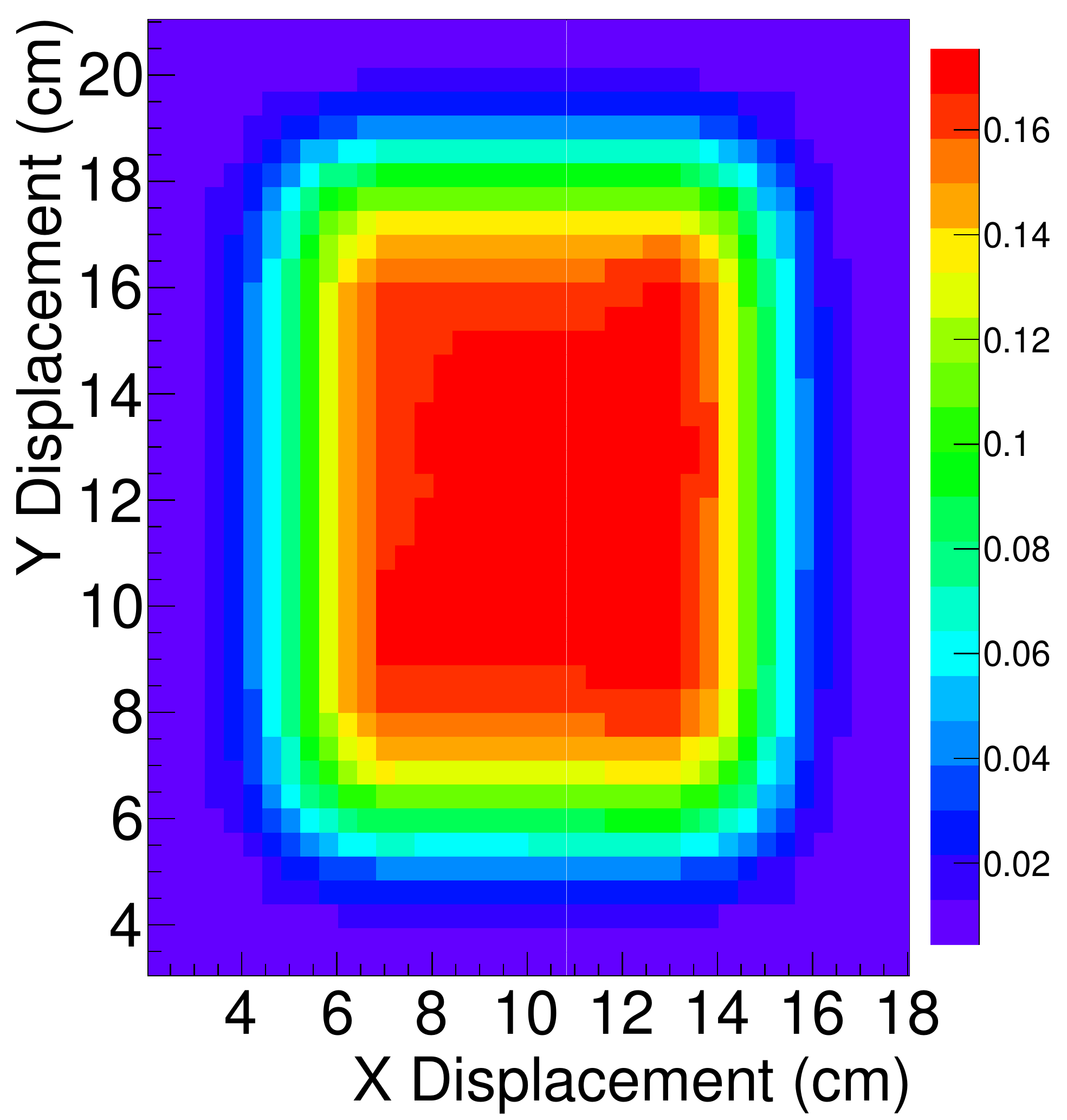}
  \caption{FnPB beam profile measured during n$^3$He comissioning in $1\,\mathrm{cm}$ steps.}
  \label{fig:FnpBBeamProfiles}
\end{figure}

The minimum required interior opening size of the wire frames was taken to be the size of the neutron beam after two meters of beam divergence in air, an estimated $14\,\mathrm{cm}\times16\,\mathrm{cm}$, from the initial $10\,\mathrm{cm}\times12\,\mathrm{cm}$ beam cross section at the exit of the beam guide.  Fig.~\ref{fig:FnpBBeamProfiles} shows a scan of the beam profile, taken during the experiment commissioning. The final internal size each frame was $16\,\mathrm{cm}\times16\,\mathrm{cm}$ and the beam was ultimately collimated to $10\,\mathrm{cm}\times10\,\mathrm{cm}$, during regular production running. The size of each cell, as given by the wire spacing, has two primary competing effects. Smaller wire cells lead to a higher directional resolution and therefore to a potentially greater sensitivity to the asymmetry. However, because the proton (and triton) have a finite track length and the signals from a large collection of events were integrated, a smaller cell size also lead to an increased correlation between cells, reducing the statistical sensitivity. Both of these effects are also influenced by the gas pressure. Simulations were performed to set the cell size and determine the optimum target gas pressure. The simulation results indicated only a small (few percent) increase in the the statistical error when varying the spacing between signal wires, between $5\,\mathrm{mm}$ and $20\,\mathrm{mm}$. The decision on the final wire spacing of $19\,\mathrm{mm}$ was made primarily based on structural and spatial issues, such as frame rigidity and mounting and feedthrough space, as well the cost associated with the front-end electronics and DAQ channels for each wire.

Structural design requirements were basically dictated by two aspects: First, the wires had to be held under tension to prevent vibrations due to microphonic pickup, adding noise to the measured signal. The wires were relatively thick ($0.5\,\mathrm{mm}$) to prevent gas gain, which is generally non-linear and would potentially introduce false asymmetries in this type of measurement. Therefore the wire tension produced significant forces on the frames. The other aspect involved accurate wire frame alignment, which is further described below. The overall size of the wire frames was set by the size of the beam size at the location of the chamber (see fig.~\ref{fig:FnpBBeamProfiles}).

Macor ceramic was chosen for the frame material, despite its high costs and difficulty in machining, primarily because of its structural strength and electrical properties.  It is sufficiently rigid to minimize frame deformation and the corresponding misalignment problems.  It has a high dielectric constant and high surface resistivity to minimize leakage current from the high voltage.  Macor is radiation hard, dimensionally stable, and maintains its mechanical properties under neutron irradiation~\cite{COGHLAN1991391}. Macor is also a low outgassing material, making it compatible with operation under high vacuum~\cite{NASAOutgassing} and limiting the production of contaminants in the chamber.  The frames were manufactured by Accuratus Corp.~\footnote{https://accuratus.com/}.

\begin{figure}[tbp]
 \centering
  \includegraphics[width=0.8\columnwidth, keepaspectratio]{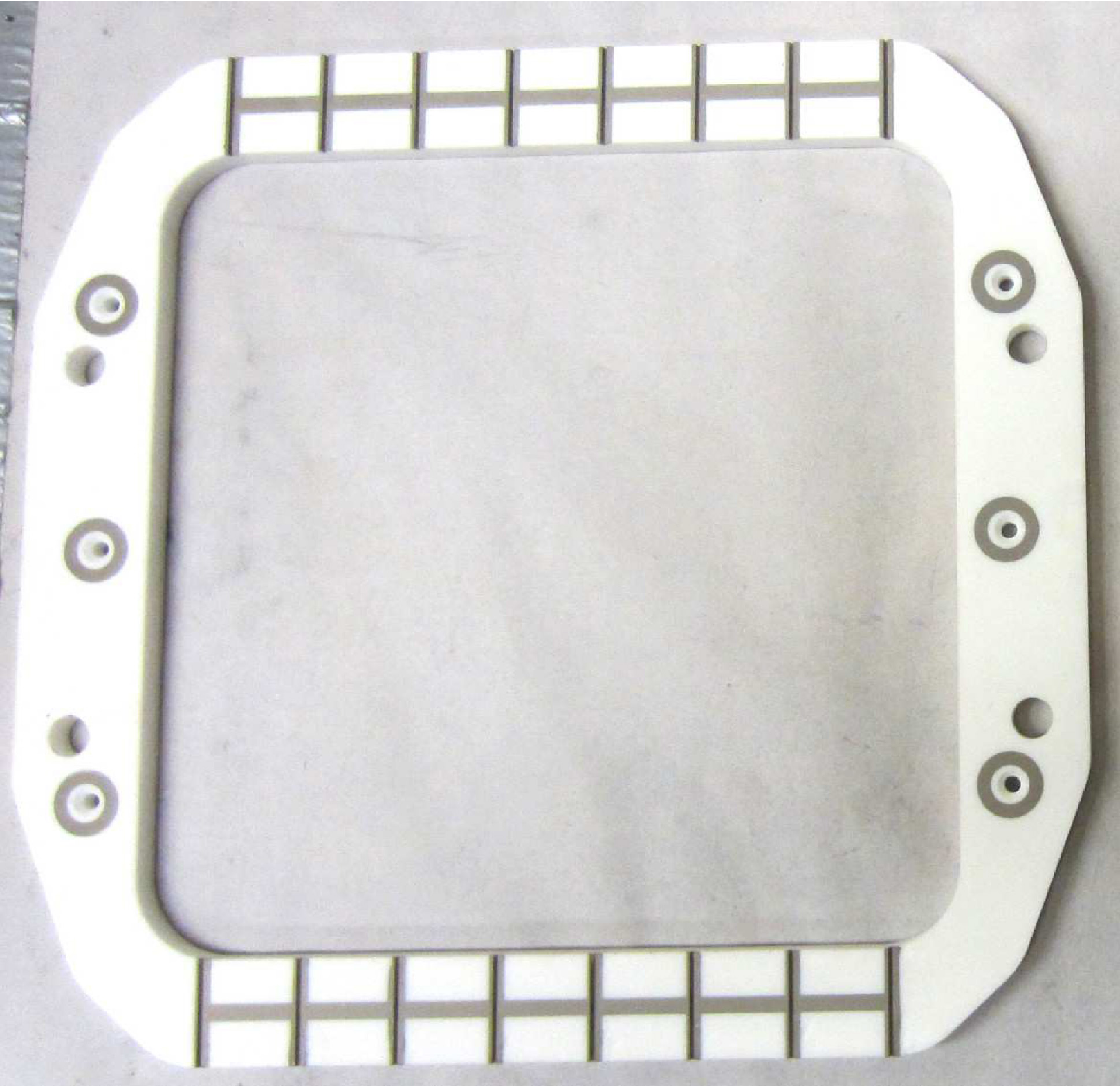}
  \caption{Macor HV frame, with conductive hick film metalization (gray). The metalized rings around the mounting holes are connected to ground to drain potential surface currents.}
  \label{fig:n3HeWireFrames}
\end{figure}

Autodesk Inventor\circledR~\footnote{https://www.autodesk.com/education/free-software/inventor-professional} single part stress simulations were performed using different frame thicknesses and wire spacings. Thinner frames would have allowed a closer wire spacing, but also require more wires per frame increasing the stress on the frame and the cost of the read-out electronics.  The selection criteria was that the safety factor should not go below $2$ in the stress analysis. Using the results of the CAD simulations, a frame thickness of $6.35\,\mathrm{mm}$ was chosen with a $19\,\mathrm{mm}$ wire spacing.  Fig.~\ref{fig:n3HeWireFrames} shows one of the HV frames without the wires. To attach the wires to the macor frames, grooves were cut into the frame surface to ensure precise, even positioning of the wires in each frame. Thick film metalization~\footnote{http://www.hybridsources.com/} was then used inside the grooves, to create solder pads for the copper wire, as well as for grounding pads.

\subsubsection{Wire Frame Alignment and Mounting}

To minimize systematic effects in the measurement, such as mixing the PV and PC asymmetries due to frame rotation and twist, one of the design goals of the n$^3$He target was to be able to align the wires in the target to about $1\,\mathrm{mrad}$ with respect to each other as well as with respect to the magnetic holding field that determines the average beam polarization. For the wire frames, this meant that machining tolerances of better than $0.25\,\mathrm{mm}$ were required in all dimensions, which was reasonably straightforward. A large PC asymmetry was measured in dedicated runs, giving $A_{PC} = (-43.7 \pm 5.9~\mathrm{(Stat)})\times 10^{-8}$. Any mixing of this asymmetry into the PV asymmetry needed to be controlled to at least an order of magnitude better than the goal statistical error on the measured PV asymmetry. To limit mixing of the asymmetries to less than $1\times10^{-9}$, the following alignment condition needed to be satisfied:
\begin{equation}
\phi < \sin^{-1}\left(\frac{10^{-9}}{10^{-6}}\right) \simeq 0.57\,\mathrm{mrad}~,
\end{equation}
where the assumption was made, based on calculational estimates prior to chamber construction~\cite{n3HeRev-2009}, that the asymmetry was about a factor of 2 larger than what was measured. For the measurement of the PV asymmetry, $\phi$ is the maximum allowed deviation from a $90^{\circ}$ angle between the wires and the magnetic holding field. This is a fairly strong requirement, but, as discussed in section \ref{sec:n3HeFiducialization}, the frame stack and assembled target was examined by the the SNS Survey and Alignment Group to fiducialize and successfully align the target.

\begin{figure}[htbp]
  \centering
  \includegraphics[width=0.45\textwidth]{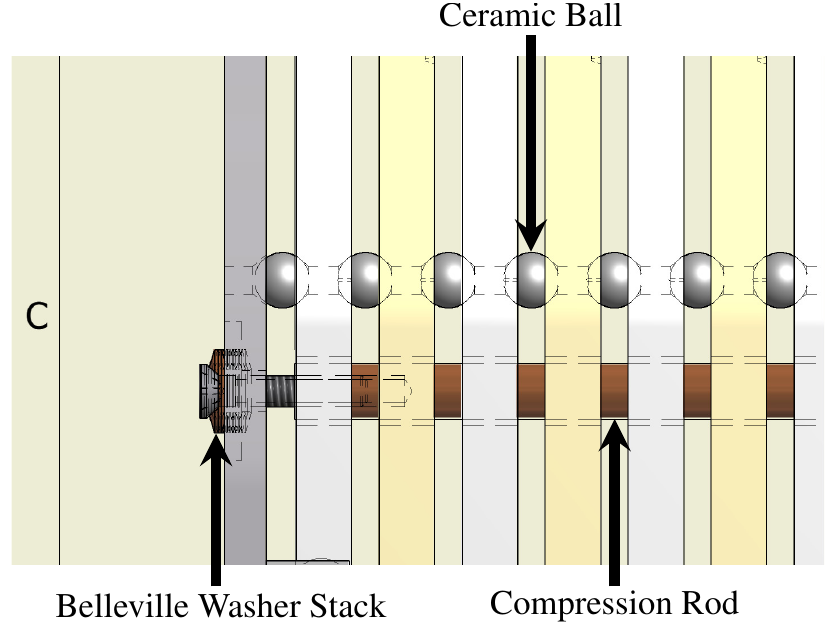}
  \caption{CAD model of the ball joint that was used between the wire frames. One of the compression rods is also visible with the screw and spring washers used to set the pre-tension in the rod. The compression plate is shown in gray.}
  \label{fig:KinematicMount}
\end{figure}

To minimize the misalignment between frames, a kinematic mount consisting of three ball and cone joints was used, since this type of mount can provide accurate and repeatable alignment between surfaces.  A set of six conical sections were machined into the surface, around the mounting rod though-holes, on each side of a frame (see fig.~\ref{fig:n3HeWireFrames}). Three of these were used in the assembly, as contact points between frames. The wire frames were stacked with a set of three ceramic beads between the frames, and this gave a reliable and consistent positioning between the frames. Fig.~\ref{fig:KinematicMount} shows a cross section of the frame stack CAD model showing the ceramic balls in the positioned between the frames.

Fig.~\ref{fig:KinematicMount} shows a CAD model detail of the frame stack assembly. To hold the frame stack together a spring loaded compression plate (gray plate in fig.~\ref{fig:KinematicMount}) was used, together with four phosphor bronze rods (brown). This also allowed careful control of the total compression force on the frames and to allow a limited range of thermal expansion and contraction in the frame stack without damaging the frames. The compression plate made contact with the frame stack through the same 3 point kinematic mounting as is used between the frames.
%additional images from appendix of thesis-> %Fig.~\ref{fig:IonChamber:CompressionPlate} shows the frame stack with the compression plate on top and fig.~\ref{fig:IonChamber:CompressionPlateSprings} shows the stack of Belleville spring washers under the bolt head that was used to set the spring pre-load.  Fig.~\ref{fig:FrameStack} shows the assembled frame stack just before insertion in the housing.

The frame stack was mounted using an aluminum plate that was friction clamped to the inside edge of the upstream window of the target chamber. The frame stack was mounted to this plate using the same kinematic mount scheme as was used between the wire frames. Four threaded holes in the mount plate were used to hold the phosphor bronze rods for the compression plate, as shown on the right, in fig.~\ref{fig:KinematicMount}. A similar plate was used at the downstream end of the frame stack (seen in fig.~\ref{fig:chamberfig}). However at that end, the plate position was adjustable, using bolts, so that the frame could be aligned and to reduce torque or twist on the frame as much as possible.

\subsubsection{Signal Routing and Wire Frame PCBs}

\begin{figure}[tbp]
  \centering
  \includegraphics[width=0.48\textwidth]{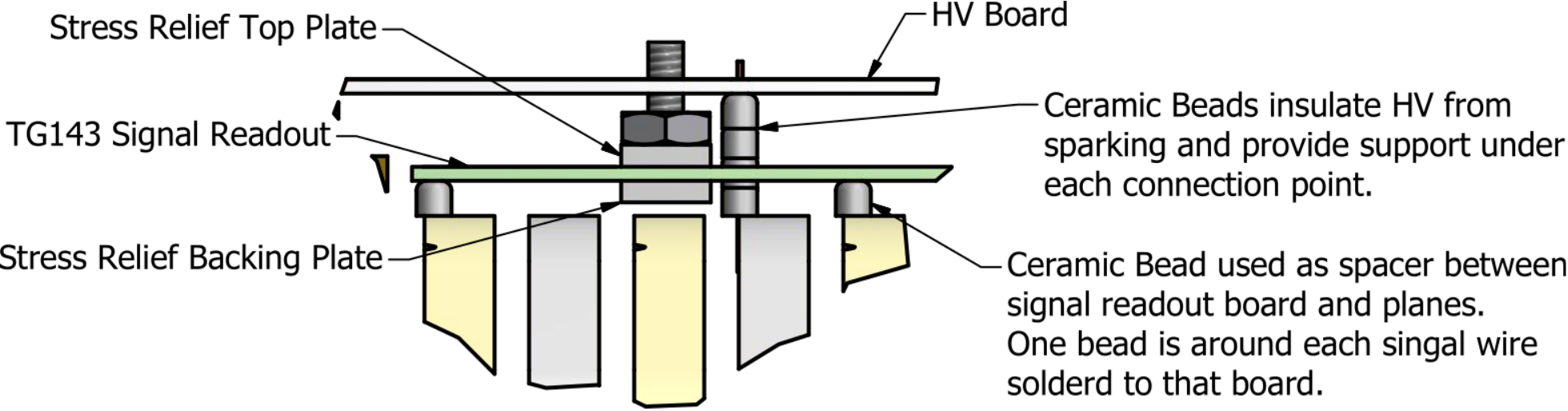}
  \caption{CAD model detail of the mounting scheme for the signal and HV PCBs, including the use of ceramic beads as spacers between the PCBs and the wire frames.  The stress relieving plates are used to clamp the Kapton ribbon cables to the signal PCBS to prevent damage to the solder joints and traces during installation.}
  \label{fig:PCBSpacers}
\end{figure}

To read out the 144 signal wires and supply the high voltage (HV) bias, two printed circuit boards (PCBs) were attached to the top and bottom of the wire frame stack as shown in fig.~\ref{fig:PCBSpacers}. These PCBs were used to route the signal and HV connections from the wires to the feedthrough ports on the chamber. The wide signal PCB and the narrow HV PCB can be seen mounted on one side of the frame stack in fig.~\ref{fig:FrameStack}. Each signal wire was soldered to a separate trace on the PCB which were then connected to 8 high vacuum capable 25 pin D-Sub feedthroughs using Kapton ribbon cables. The HV wires were simply multiplexed to two standard SHV feedthroughs, which were connected to the HV PCBs with copper wire, isolated with ceramic beads.

The PCBs were made from Rogers' Duroid material \footnote{https://www.rogerscorp.com/acs/producttypes/6/RT-duroid-Laminates.aspx}, which is suitable for use in UHV. Since Duroid is a Teflon glass composite, it is also fairly radiation hard. The main effects of radiation exposure are an increase in brittleness and a reduction in tensile strength.  The boards were separated from the HV planes with ceramic bead spacers, placed over each signal wire, which also served to shield the signal wires from the HV.

% as shown in fig.~\ref{fig:ReadoutCabling}.
%\begin{figure}[htbp]
%    \centering
%    \includegraphics[width=0.45\textwidth, height=0.3\textheight, keepaspectratio]{figures/IonChamber-Cabling-EvenPlanes}
%  \caption{Signal Read out PCBs for the odd numbered wire frames.  Each signal PCB carried one half of the signal frames.}
%  \label{fig:ReadoutCabling}
%\end{figure}

% as shown in fig.~\ref{fig:PCBSpacers}.

\subsection{Chamber Fiducialization}\label{sec:n3HeFiducialization}

To reach the required alignment sensitivity, the frame stack was fiducialized by the SNS Survey and Alignment Group. 3D coordinate measurements of the frame stack position were taken relative to the chamber mount flange and four fiducial markers that were glued on the exterior of the flange (one of them can be identified as the black marker, seen on top of the flange at the far end in fig.~\ref{fig:chamberfig}). In this process, the outside edge of each wire frame was surveyed to measure its angular alignment and position with respect to the outside of the chamber. The wire frames were machined to a precision of $0.25~\mathrm{mm}$ in all linear dimensions, so that the maximum angular misalignment error of each wire with respect to its frame was $\sim \pm 0.5/160 \sim \pm 3~\mathrm{mrad}$ (the $160~\mathrm{mm}$ being the length of the wire). During the survey of the frame position with respect to the chamber, it was found that the frame stack had an approximate $25\,\mathrm{mrad}$ twist from the first frame to the last frame (in beam direction). This twist had to be taken into account in the final analysis of the measured asymmetries, since it mixes the PV and PC asymmetries, as can be understood from the discussion, following eqn.~\ref{eq:DiffCrossSection}. Since the statistics are highest in wire planes closest to the entrance windows, where most neutrons will capture, it was preferable to minimize the asymmetry mixing at the front of the target chamber by aligning the first wire plane to the magnetic field. All asymmetries measured in the downstream frames were adjusted with a value corresponding to the rotation of a given frame, multiplied by the measured PC asymmetry.

\begin{figure}[tbp]
  \centering
  \includegraphics[width=0.4\textwidth, keepaspectratio]{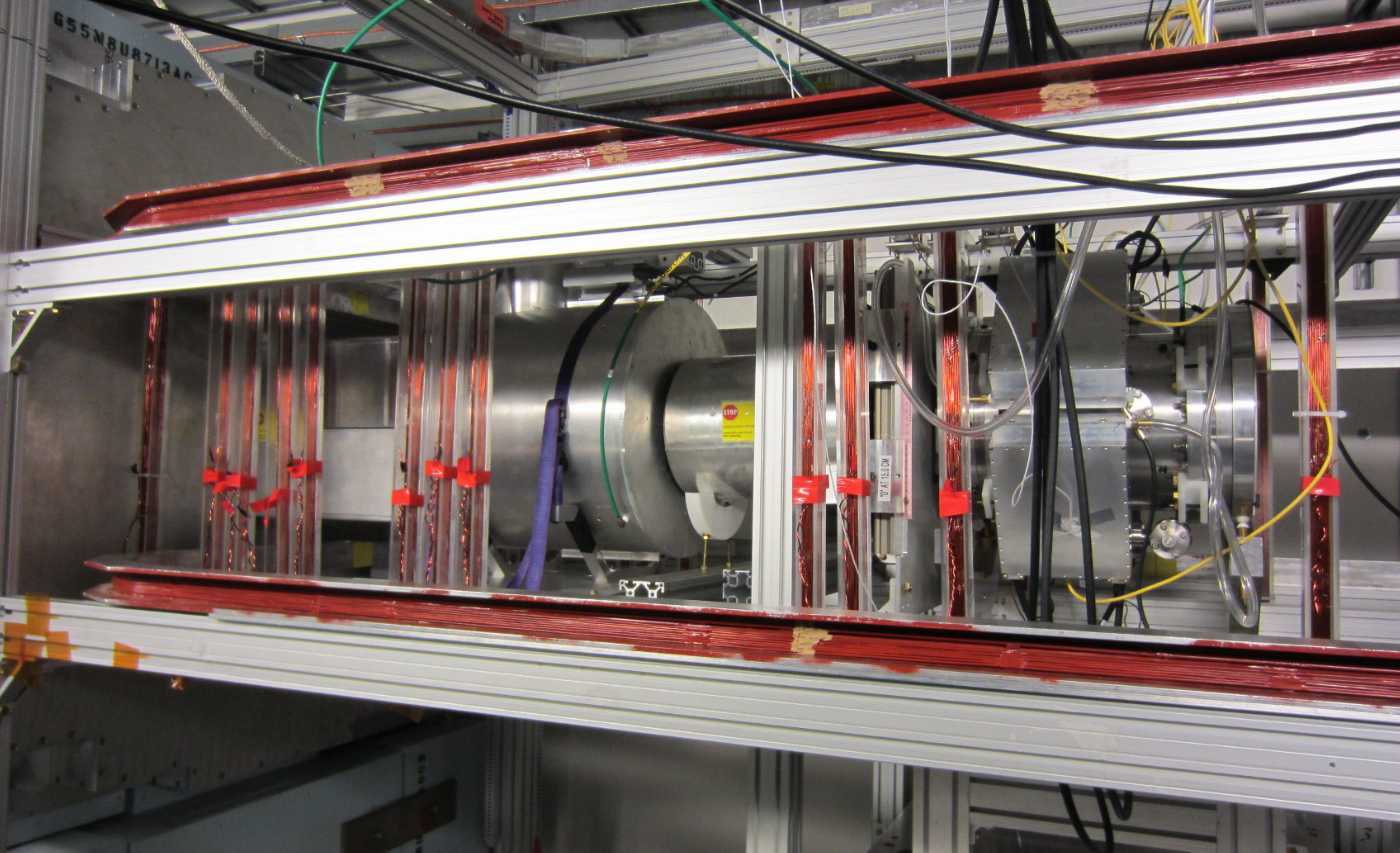}
  \caption{Fully installed experiment. From left to right: Supermirror polarizer; RF spin rotator; active target chamber. The chamber can be seen positioned on the alignment mount (to the bottom right).}
  \label{fig:FullInstall}
\end{figure}

To determine the overall chamber orientation with respect to the holding field, the alignment equipment had to simultaneously see at least three fiducial markers on the outside of the chamber, with the chamber installed on the beamline.  To this end, additional fiducial markers were glued to the exterior of the housing and located relative to the initial four reference points, before installation on the beamline. A flat level block was glued to the top of the chamber, aligned parallel to the first wire frame to serve as an external reference surface.  This level block was used with a digital protractor to make an initial rough alignment with the holding field after installation and after a $90^{\circ}$ rotation between the UD and LR measurement modes. The survey and alignment crew then made the final alignment. To make precise alignment of the target chamber possible, a four point adjustable stand was constructed, to mount the target chamber in the neutron beam. The overall uncertainty in the chamber alignment to the holding field is $\sim \pm 3~\mathrm{mrad}$, corresponding to the alignment equipment and field probe uncertainty. The compound effect of a maximum wire-to-frame and frame-to-field misalignment is therefore $\sim \pm 6~\mathrm{mrad}$, which when combined with the PC asymmetry, gives a maximum uncertainty due to PV-PC mixing, of $\delta_{A_{PV}} \pm 2.6\times10^{-9}$. This was the single largest systematic error in the measurement. The fully installed experiment can be seen in fig.~\ref{fig:FullInstall}.

%\section{Chamber Assembly}\label{sec:ChamberAssembly}
\subsection{Chamber Filling and Benchtop Testing}\label{sec:BenchTopTesting}

As mentioned in sec~\ref{sec:FrameStackDesign}, the fill gas pressure effects the asymmetry sensitivity of the chamber.  A higher pressure will shift the capture distribution more to the front of the chamber and capture more neutrons and therefore increase the statistical accuracy of the measurement. At the same time though, the concentration of events into fewer cells, at the front of the chamber, will increase the correlation between the majority of events, reducing the statistical sensitivity. On the other hand, reducing the gas pressure produces a more uniform event distribution in the chamber, but will also cause neutrons to exit the back of the chamber without detection and increase the proton mean free path, causing them to leave the active volume of the target before depositing the majority of their energy. Taking all of these effects into account, the optimal pressure was found using Monte Carlo simulations, looking for a minimum in the error dilution (the statistical error above neutron counting statistics) as a function of pressure, cell size, and energy deposition of the proton and triton (see fig.~\ref{fig:n3HeBraggCurve}). The simulation showed a broad minimum between $0.4$ and $0.5\,\mathrm{atm}$.  Based on these simulations and the HV requirements (discussed below), a gas pressure of $0.47\,\mathrm{atm}$ was chosen to allow sufficiently high bias voltage for efficient charge collection. This produced an intrinsic error dilution of about $3.5$, giving an expected statistical error of about $\sigma_{stat} \simeq 3.5/\sqrt{n}$, where is $n$ is the number of neutron captures.

\section{Simulations}\label{sec:Sims}

Simulations were used in the design process as well as in the determination of the geometry factors, which are necessary for the extraction of the physics asymmetry (see Sec.~\ref{scn:geof} below). Geant4~\cite{Geant4} was used primarily to model the neutron physics, the energy deposition in the chamber, and to study backgrounds from neutron capture on materials other than $^{3}He$. Using the neutron and electromagnetic physics implemented in Geant4 the simulations were used to find the optimum cell size and target gas pressure and the geometry factors. The sensitivity of the chamber relies on a cell design, as determined by the arrangement of bias voltage and signal wires, such that the ionization electrons produced in a given cell, by a given track, are collected mostly in that cell, with little to no signal leakage into the neighboring cells. This aspect of the chamber was studied with Garfield++~\cite{GarfieldUserGuide} simulations, including the correlation between wire cells, due to charge sharing. Another effect that produces correlation between cells is proton and triton track sharing between cells. In this case, we were able to determine this correlation from the data itself, as well as from simulation, as discussed in section~\ref{scn:comm}.

\subsection{Charge Collection Simulations}
\label{ch:GarfSim}
To ensure that each pulse was a statistically independent measurement of the physics signal it was required that all ions and electrons produced in one pulse were collected before data taking for the next pulse began.  As described earlier and shown in fig.~\ref{fig:TwoPulsePreampOut}, in each pulse, the wire signals are sampled at $20\,\mathrm{\mu s}$ intervals and summed into 49, $0.32\,\mathrm{ms}$ wide time bins. However, for the calculation of the asymmetries for each wire cell (or pair of cells), the signal for each pulse was integrated from bin 5 to 44 (inclusive), a $13.12\,\mathrm{ms}$ time frame. Together with the $0.98\,\mathrm{ms}$ gap between pulses, this meant that there was a $3.54\,\mathrm{ms}$ period, during which no data was taken. It was therefore required that the charge collection time for the ions and electrons be $\lesssim 3\,\mathrm{ms}$. In addition, both, for the purposes of chamber design and later verification of chamber performance, the simulations were used to predict the amplitude of the current that is induced on the signal wires, by the charge motion at various points in the target chamber. Finally, the simulation was also used to study cross-talk between adjacent cells, and optimize the design if the cross-talk was found to be excessive.

\subsubsection{Simulation Setup}\label{scn:GARFSETUP}

Garfield++ was identified as the most suitable simulation framework for these tasks. It was used to simulate the charge collection times and induced currents in a reduced geometry model~\cite{GarfieldUserGuide} (see section~\ref{scn:GARFSETUP} below).  The charge collection times were examined for both $^3$He$^+$ ions and electrons, starting as pairs in different positions within the target chamber, and the charge collection times and the total induced current were recorded. A 3D chamber model was used in the charge collection simulation.  Since Garfield++ cannot calculate arbitrary 3D fields, Gmsh ~\cite{gmsh,GmshWebsite} was used to define the simulation geometry and create a tetrahedral mesh. A reduced geometry model was constructed, which included the inside surface of the target chamber and all of the wires in the first 6 HV frames and 5 signal frames.  The ceramic wire frames, circuit boards, feedthroughs, and other construction details were omitted from this model.  This reduced the computation complexity of the model, decreasing the processing times when producing the field maps, and the memory requirements during charged particle drift simulations.  This wire subset allowed the charge to be simulated over three complete planes, consisting of four HV frames and three signal frames, to investigate cross talk between the cells. The Gmsh output mesh was processed using Elmer~\cite{ElmerWebsite}, an open source multi-physics simulation engine that was used to calculate the electric field for the given mesh geometry. In addition to the mesh file, Elmer required a set of boundary conditions on the mesh surfaces, and the electric permittivity of the mesh volumes, to perform the calculation. In the simulation, the housing cylinder and signal wires were grounded, while the HV wire surfaces were set at $-350\,\mathrm{V}$, close to the measured $-353\,\mathrm{V}$ bias voltage that was used during data taking. In Garfield++ gas properties are calculated using Magboltz~\cite{MagBoltzWebsite}, a program that calculates the electron transport properties in gas mixtures, using a semi-classical Monte Carlo simulation.  Ion mobilities are loaded from a user provided text file. Information for the $^3$He ion mobilities was taken from~\cite{Ellis1976177}.

\subsubsection{Simulation Results}
\begin{figure}[tbp]
  \centering
    \includegraphics[width=0.45\textwidth, height=0.3\textheight, keepaspectratio]{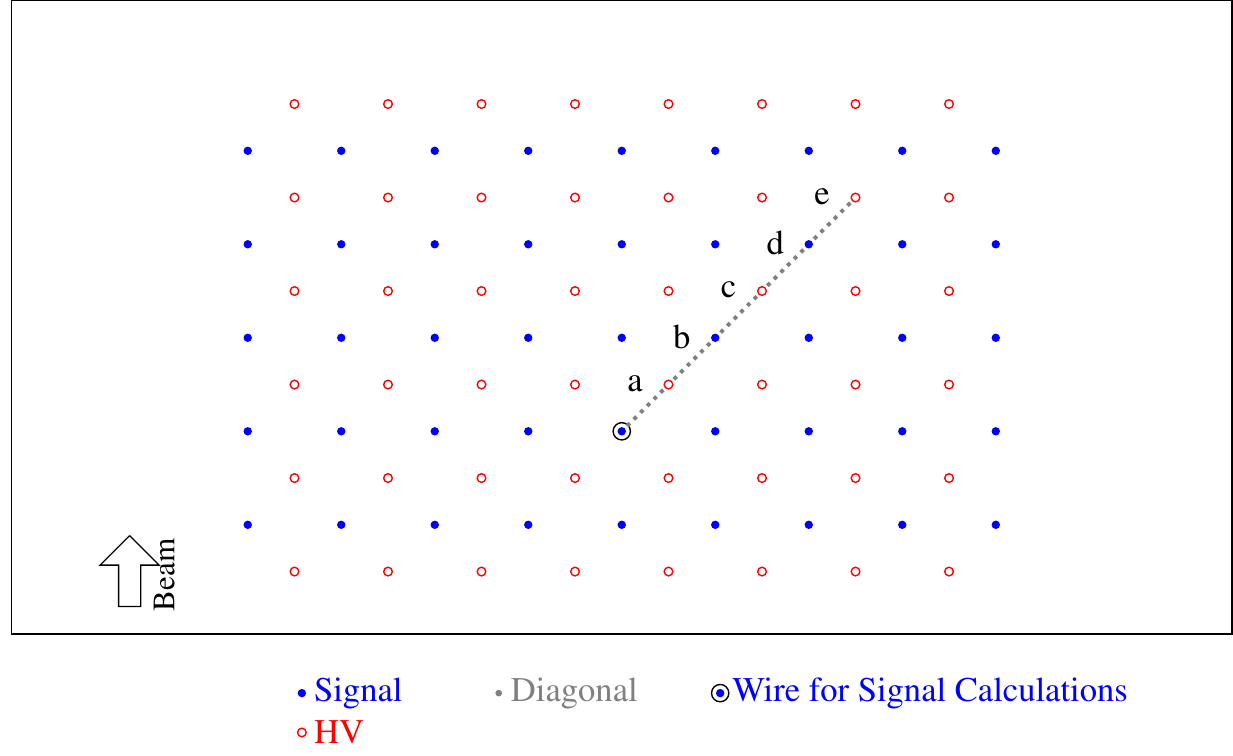}
  \caption{Location of ion-electron pair origins for crosstalk and collection time simulations. The corresponding induced charge was observed  for the signal wire marked with the black circle.  Each half cell that the simulation lines pass through is labeled alphabetically.}
  \label{fig:Garf:Layout}
\end{figure}

To examine the charge collection time and the total induced signal from ionization within a cell and from adjacent cells in the target chamber, a paired ion and electron were drifted 100 times from multiple locations in the model.  For illustration, a set of thirty of the starting locations equally spaced along a diagonal line passing through the corners of the cells as shown in fig.~\ref{fig:Garf:Layout} is discussed in more detail.  Each half cell that was crossed is marked with a lower case letter from ``a'' to ``e''.  For all starting locations the induced signal current is calculated for the signal wire marked with the black circle at the lower left of the line.  By calculating the signal induced on this one wire, from charge drift starting at all points, the level of crosstalk between the cells can be calculated.

The longest simulated electron collection times were of order $2\,\mathrm{\mu s}$, and the longest simulated ion collection times were of order $1\,\mathrm{ms}$, which is well below the required maximum of $3\,\mathrm{ms}$ to keep the neutron pulse signals independent.

The current induced on one of the grounded signal wires by the motion of the ionization charges can be calculated using the Shockley-Ramo theorem~\cite{SpielerSemiconductor}. In general, if a particle of charge $q$ travels a path starting at the surface of one conductor and ending on the surface another conductor then the total charge collected is equal to the integrated current over the collection time and will be $-q$ on the starting conductor, and $q$ on the ending conductor.  If a particle of charge $q$ starts at some arbitrary point along the path and a second particle of charge $-q$ starts at the same point but moves in the opposite direction along the path, then we expect the total induced currents on the conductors to be the same as one particle traversing the entire length of the path when the particles of opposite charge reach the ends of the path.  For conductors that do not form a start or termination point for a given simulated path, a bi-modal current will be induced, that will integrate to zero over the full collection time.

\begin{figure}[h]
\includegraphics[width=0.48\textwidth]{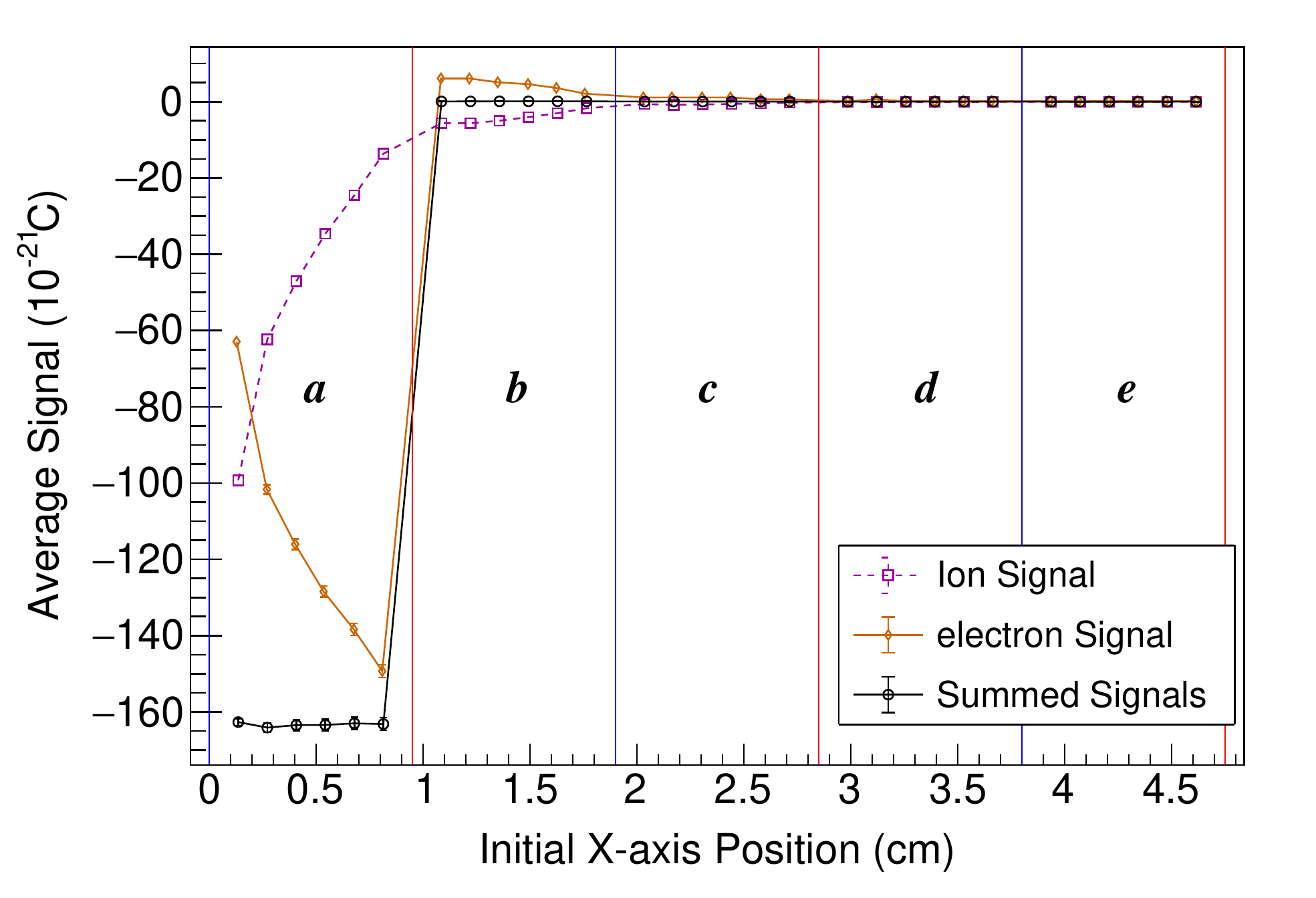}
\caption{The mean integrated currents for electrons, ions, and the summed ion and electron signal, with standard deviation error bars from 100 simulated events for the starting locations shown in figure \ref{fig:Garf:Layout} (with the letter indicating the same half cells).  Each red line shows where an HV wire is crossed and each blue line shows where a signal wire is crossed. The first cell $(0 < x  < 1)$ is the signal cell and the remaining cells are the neighboring cells with essentially zero charge collection.}
\label{fig:IntCharge}
\end{figure}

The mean and standard deviation of the integrated currents was calculated for the 100 ion and electron events, starting at each of the given locations and is plotted in fig.~\ref{fig:IntCharge}.

As can be seen in~\ref{fig:IntCharge} the maximum value of cross talk from the bimodal signal is roughly $6\times10^{-21}\;\mathrm{C}$ in cell ``b'', and it rapidly drops off in more distant half cells. This corresponds to about $4\%$ of the in-cell single event signal, but as we are integrating the current signal over $13.12\;\mathrm{ms}$ most of the cross talk will integrate to zero.  With the ion collection times approximately linearly distributed from  $0 \rightarrow 1\;\mathrm{ms}$, it can be assumed that half of the ion signals are fully integrated in that millisecond, and will cancel half of the electron signal, and similarly for the non-diagonal adjacent cells with ion and electron signals of order $25\times10^{-21}\;\mathrm{C}$. Assuming a constant rate of ionization over the integration time then $1\;\mathrm{ms}/13.12\;\mathrm{ms} = 7.6\%$ of ionization occurs in this time period. The overall result is that, for any given wire cell, the fraction of the total signal seen in that cell, due to cross-talk is only about $0.027$.  One of the assumptions in the calculation of the geometry factors is, that all of the signal in each cell arose from ionization only in that cell and the contribution from cross-talk has a negligible effect on the geometry factors. See section \ref{scn:geof} for details on the geometry factor calculation.

\subsection{Geometry Factors}~\label{scn:geof}

A determination of the target-detector geometry factors at the $\leq 10\%$ level is crucial for the extraction of the physics asymmetry from the measured (uncorrected) asymmetry. Equation~\ref{eq:wireyield} shows the dependence of the wire yield on the geometry factors, which essentially quantify the acceptance of each wire cell. Since the detector chamber was operated in integration mode, these factors correspond to an energy deposition weighted average over the track angles for all possible proton and triton tracks through a cell, for each neutron capture within a single beam pulse. For example, the geometry factor for the $\cos\theta$ angle in eqn.~\ref{eq:DiffCrossSection} for cell $i$ is given by
\begin{equation}
G^{PV}_{UD_i} = \frac{\sum\limits^{N}_{k=1} \left(E^p_i +  E^T_i\right)_k\cos{\theta_{k,i}}}{\sum\limits^N_{k=1} \left(E^p_i +  E^T_i\right)_k}~.\label{eqn:geof}
\end{equation}
Where $N$ is the number of neutron captures in the active volume of the target, $E^{p,T}_i$ are the proton and triton energy depositions in cell $i$, which follow the curves in fig.~\ref{fig:n3HeBraggCurve}, and $\theta_{k,i}$ is the angle of the proton track, measured with respect to the holding field direction, for track $k$, going through cell $i$.
\begin{figure}[h]
  \centering
    \includegraphics[width=0.45\textwidth]{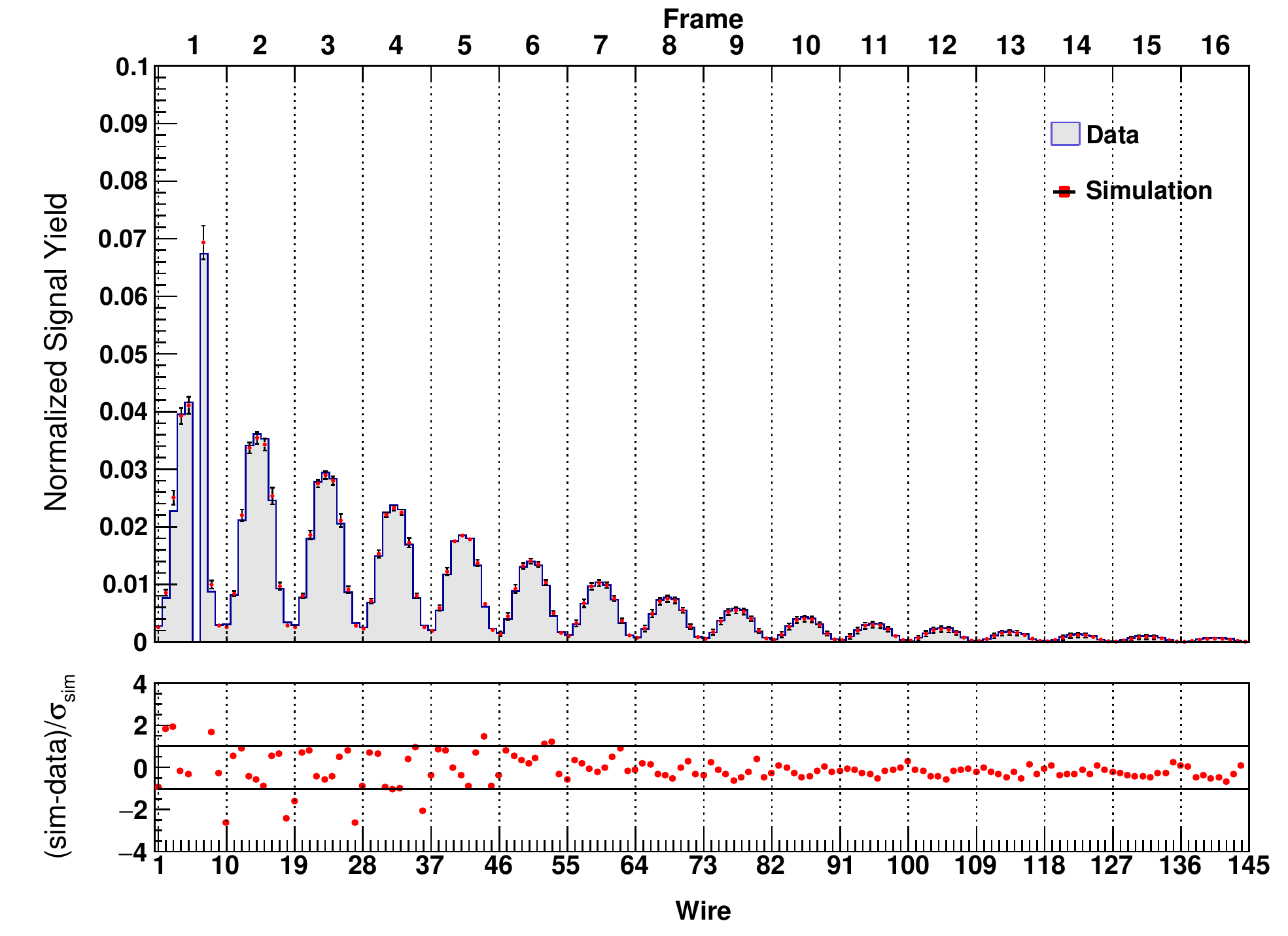}
  \caption{Comparison of GEANT4 simulation results to the wire yield, measured during a run in with standard (production mode) beam collimation. In the top graph, the shaded histogram corresponds to the measured data in each wire, normalized to the integrated chamber signal, while the red data points correspond to the simulated wire response normalized to the total number of events in the chamber. The bottom plot shows the standard deviation between the simulated and measured normalized yields, with respect to the simulation error. The error bars on the measured data are too small to be visible on the plot and the errors on the simulated data are a result of varying both, the position of the chamber relative to the beam centroid and the fill gas pressure, within the range of measurement uncertainty ($\pm 3$~mm and $0.46 - 0.49$~atm). These measurements were taken in the PV mode, with the wires running int the horizontal direction. Each peak in the top plot corresponds to a wire frame and the horizontal axis indicates the continuously increasing wire number, where the lowest wire number in each frame corresponds to the bottom wire in the frame. Wires 6 and 7 were accidentally shorted and excluded from the data analysis (the simulation yield was for the two wires was summed). Several of these comparisons were done, with the beam collimated to different regions of the chamber.}
  \label{fig:BeamScanRunPlot}
\end{figure}

\begin{figure*}[ht]
  \centering
    \includegraphics[width=0.9\textwidth]{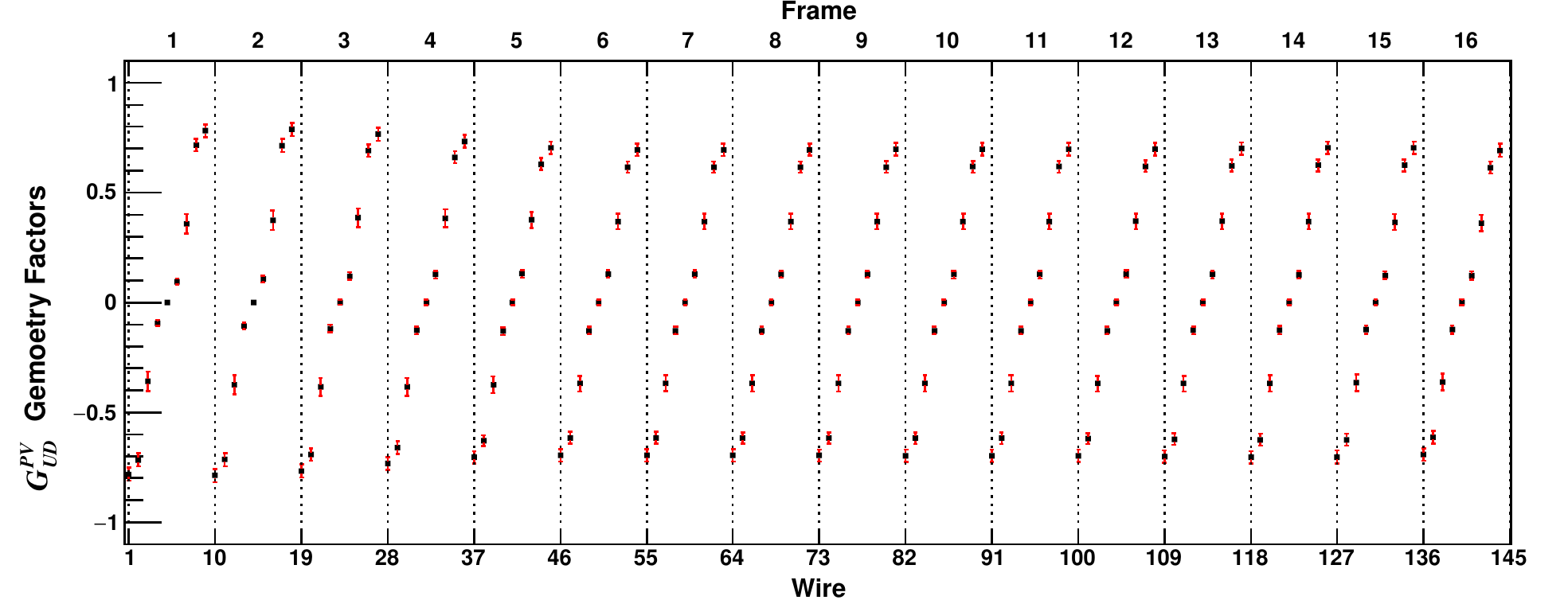}
  \caption{Shown here are the UD-mode (wire orientation is horizontal) PV geometry factors replacing $\cos\theta_y$ in in eqn.~\ref{eq:DiffCrossSection} (see eqn.~\ref{eqn:geof}). The error bars are from systematic variations in the simulation, over chamber position and target pressure, within measurement uncertainties.}
  \label{fig:PVUDGeoFactors}
\end{figure*}

The geometry factors are calculated using eqn.~\ref{eqn:geof}, by simulating many proton and triton tracks from neutron captures with the as-built target geometry and fill pressure. Figures~\ref{fig:nCapDist} and~\ref{fig:n3HeBraggCurve} illustrate the benchmarking that was done to verify that the correct physics behavior was implemented in the simulation. The geometry factors that are obtained with simulations are highly dependent on the implemented geometry and physical properties, which included the lateral chamber position with respect to the beam center (in $\hat{x}$ and $\hat{y}$), chamber rotations, and target fill pressure. To investigate to what degree the simulation geometry agreed with the physical geometry, the normalized simulated wire yields were compared to the normalized measured wire yields for a variety of collimation conditions. One of the comparisons between simulation and data for a run with production mode beam collimation is shown in fig.~\ref{fig:BeamScanRunPlot}. Production mode here refers to data taken in the UD mode (PV asymmetry measurement), with the wires oriented in the horizontal direction and with $10~\mathrm{cm} \times 12~\mathrm{cm}$ beam size.
\begin{figure*}[ht]
  \centering
    \includegraphics[width=0.9\textwidth]{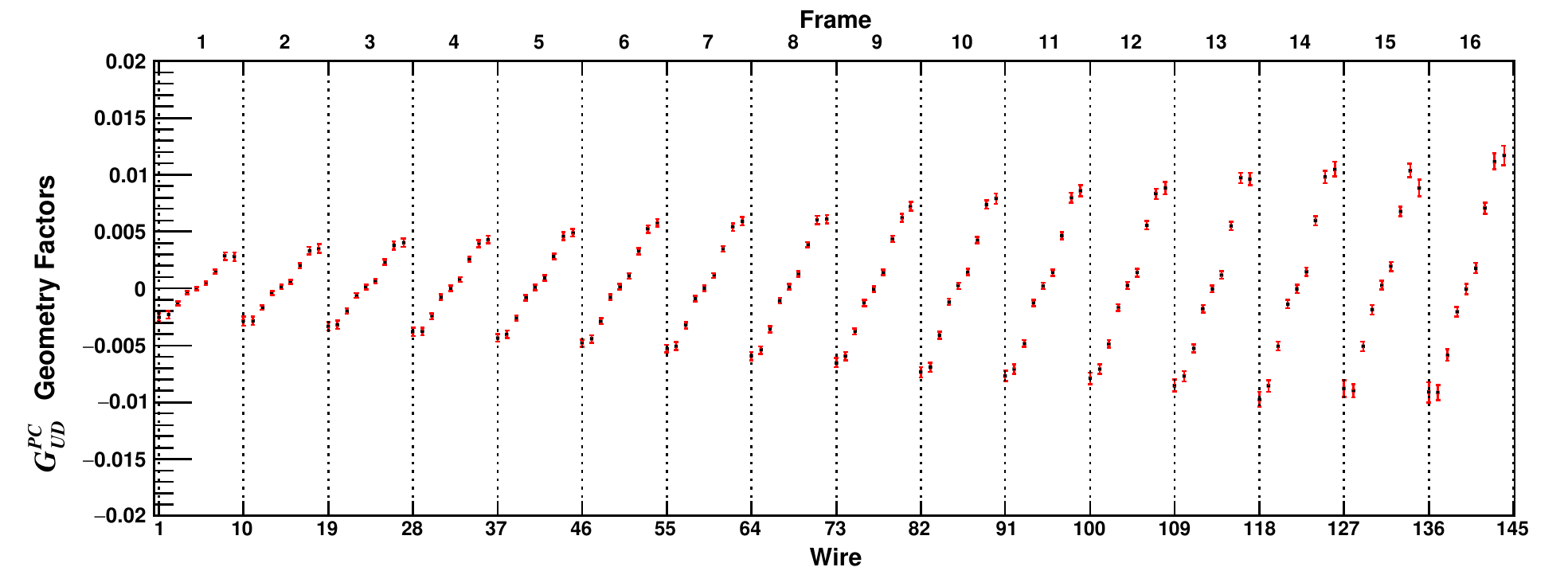}
  \caption{Shown here are the UD-mode (wire orientation is horizontal) PC geometry factors replacing $\cos\theta_x$ in in eqn.~\ref{eq:DiffCrossSection}. The error bars are from systematic variations in the simulation, over chamber position and target pressure, within measurement uncertainties.}
  \label{fig:PVLRGeoFactors}
\end{figure*}

The top plot in fig.~\ref{fig:BeamScanRunPlot} shows 16 peaks, corresponding to the 16 wire frames. Within each frame, the wires are numbered starting with the lowest number corresponding to the bottom wire, in increasing order (see figs.~\ref{fig:chamberfig} and~\ref{fig:wirelayout}). The measured yield from data runs, for each signal wire was normalized to the total yield in the chamber. Likewise, the simulated wire yield was normalized to the total number of simulated neutron captures in the target. The relative yield is highest for the center wire (in the vertical direction) in each frame and falls off toward the bottom and top wires located at the upper and lower edges of the chamber active volume.

The chamber axis (along the beam direction) was aligned with the beam centroid to within a few millimeters (which was established from beam profile measurements and installation survey). The errors on the simulated data are not from simulation statistics (which were too small to be visible). Rather, what can be seen are systematic simulation errors, which are a result of varying the position of the chamber relative to the beam centroid and the fill gas pressure, within the range of measurement uncertainty ($\pm 3$~mm and $0.46 - 0.49$~atm respectively), as well as by rotating the chamber in all directions (pitch, roll, yaw) with respect to the holding field, within the measurement uncertainty ($\pm 3$~mrad). The error on the measured yield is too small to be visible in the histogram. The bottom plot shows the standard deviation between the simulated and measured relative wire yield, with respect to the simulation error. The first wire frame shows the most significant deviations. Since this is the frame closest to the beam entrance window of the chamber, we expect this difference to come primarily from incomplete charge collection, due to field inhomogeneities. Wires 6 and 7 malfunctioned and were excluded from all aspects of the data analysis.
\begin{figure}[ht]
  \centering
    \includegraphics[width=0.9\columnwidth]{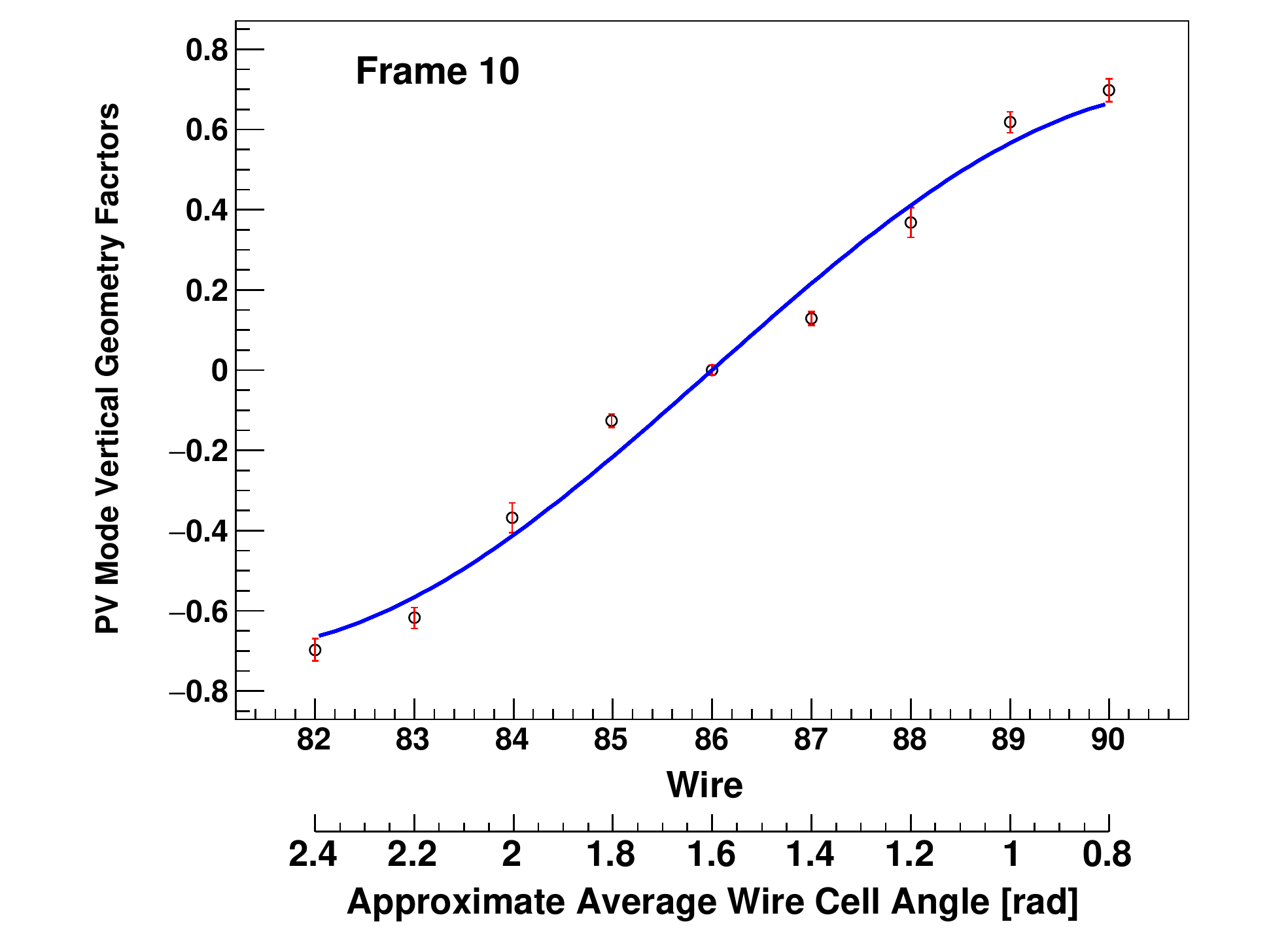}
  \caption{PV-mode geometry factors replacing $\cos\theta$ in in eqn.~\ref{eq:DiffCrossSection} for frame 10. The blue line illustrates the cosine dependence; it is not a fit to the data.}
  \label{fig:PVUDGeoFFit}
\end{figure}
The same variation in physical chamber properties that produced the errors shown in fig.~\ref{fig:BeamScanRunPlot} were used to establish the geometry factor errors shown in figs.~\ref{fig:PVUDGeoFactors},~\ref{fig:PVLRGeoFactors}, and~\ref{fig:PVUDGeoFFit}.

Figure~\ref{fig:PVUDGeoFactors} shows the PV geometry factors, corresponding to $\cos\theta_y$ in eqn.~\ref{eq:DiffCrossSection},
while fig.~\ref{fig:PVLRGeoFactors} shows the PC geometry factors corresponding to $\cos\theta_x$ in eqn.~\ref{eq:DiffCrossSection}. Figure~\ref{fig:PVUDGeoFFit} illustrates the cosine dependence for wire frame 10 (the blue line is not a fit to the data). The shown geometry factors were simulated with the target chamber in UD mode configuration, with the wires oriented horizontally, so the PV geometry
factors in fig.~\ref{fig:PVUDGeoFactors} are large, while the PC geometry factors in fig.~\ref{fig:PVLRGeoFactors} are small (note the different scale on the vertical axis).

\begin{figure}[h]
  \centering
    \includegraphics[width=0.9\columnwidth]{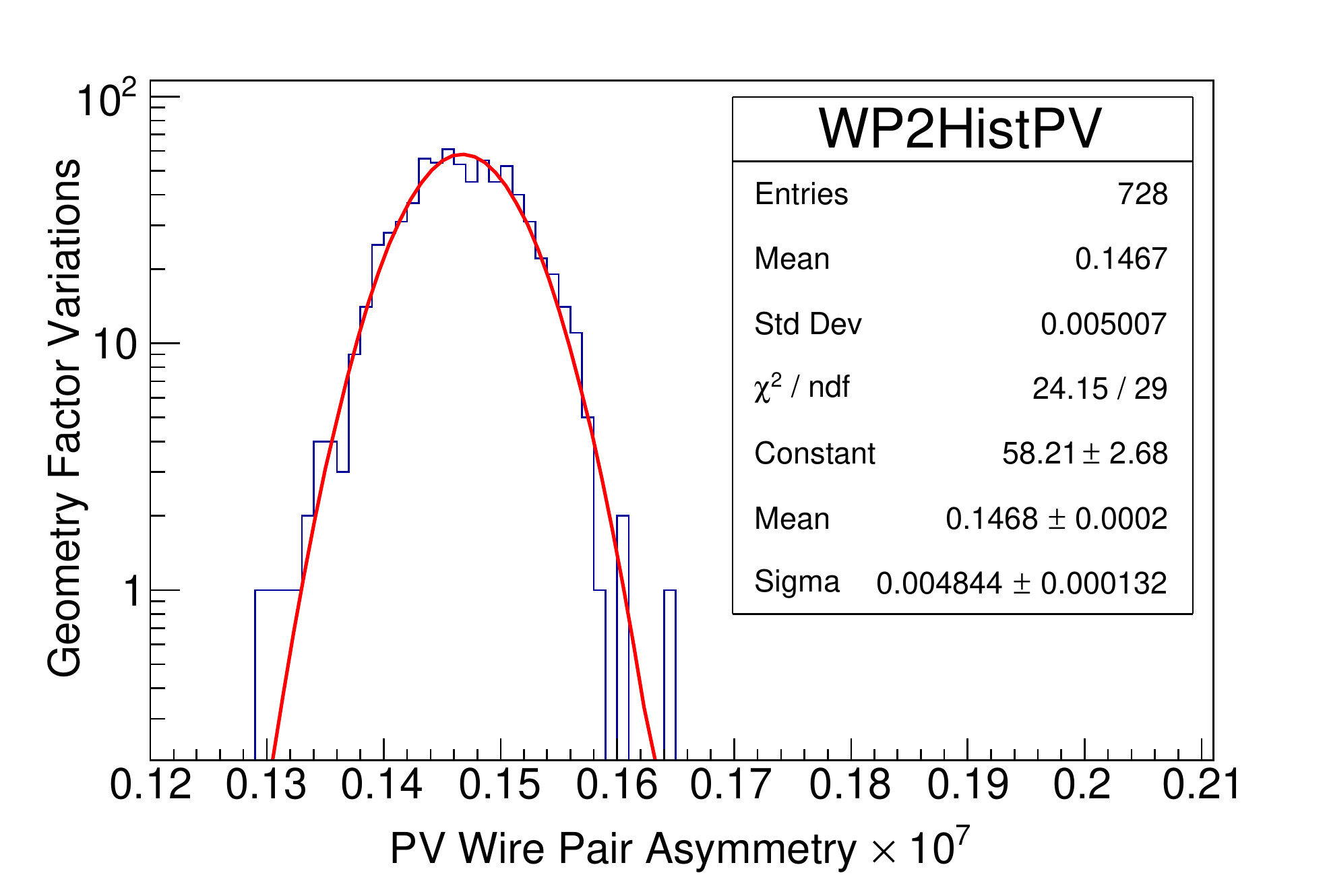}
  \caption{Distribution of wire pair PV asymmetries that were obtained as a result of varying the geometry factor values in eqn.~\ref{eqn:LeastSq} within the same range that established the simulation to yield agreement in Fig.~\ref{fig:BeamScanRunPlot}. }
  \label{fig:PVAsymGeoVar}
\end{figure}

\begin{figure}[h]
  \centering
    \includegraphics[width=0.9\columnwidth]{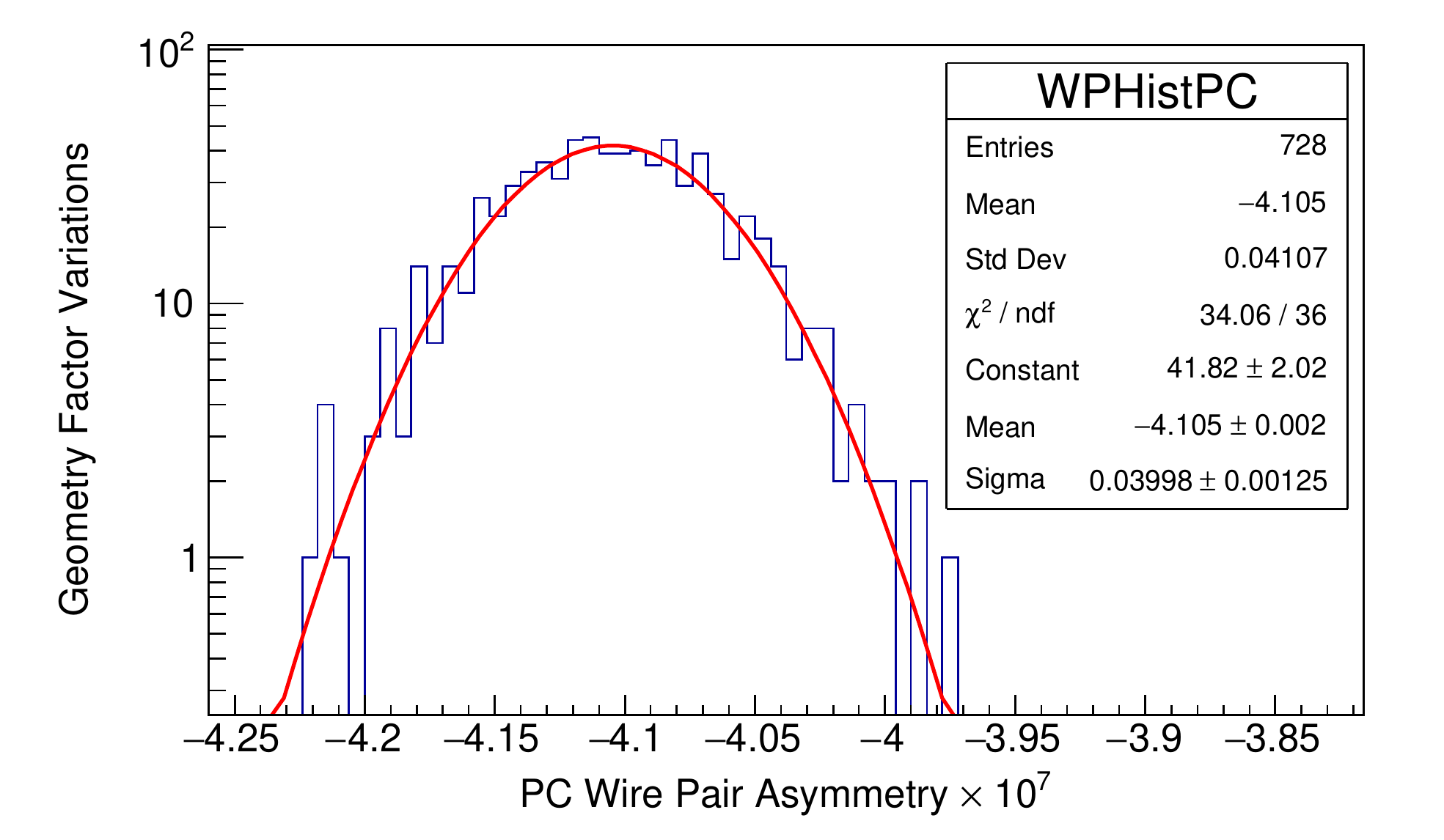}
  \caption{Distribution of wire pair PC asymmetries that were obtained as a result of varying the geometry factor values in eqn.~\ref{eqn:LeastSq} within the same range that established the simulation to yield agreement in Fig.~\ref{fig:BeamScanRunPlot}. }
  \label{fig:PCAsymGeoVar}
\end{figure}

If the chamber had been perfectly aligned with the beam and the wire rotation with respect to the holding field had been exactly $90^\circ$, for each frame, the PC geometry factors should be zero in UD mode. However, the slight offset of the chamber with respect to the beam and the frame twist, discussed in sec.~\ref{sec:n3HeFiducialization}, introduces the small but finite horizontal geometry factors shown in fig~\ref{fig:PVLRGeoFactors}, which then cause a small amount of mixing of the PC asymmetry into the PV asymmetry. The same situation is also present in LR mode running, where a small amount of PV asymmetry is mixed into the dominant PC asymmetry. The combined effect is shown in eqn.~\ref{eqn:LeastSq}.
\begin{eqnarray}\label{eqn:LeastSq} \nonumber
  A^{meas}_{UD,i} & = & \epsilon P \left( A_{PV} G^{PV}_{UD,i} + A_{PC} G^{PC}_{UD,i} \right) \\
  A^{meas}_{LR,i} & = & \epsilon P \left( A_{PV} G^{PV}_{LR,i} + A^{PC} G^{PC}_{LR,i} \right)~.
\end{eqnarray}
To extract the physics asymmetries and establish the error corresponding to the uncertainty in the geometry factors, a least-squares minimization of eqns.~\ref{eqn:LeastSq} was performed for each set of wire geometry factors that were obtained in the process of matching data and simulation depicted in fig.~\ref{fig:BeamScanRunPlot} (see also sec.~\ref{scn:CorrAna}). The final results for the variation in the geometry factors are shown in fig.~\ref{fig:PVAsymGeoVar} for the PV asymmetry and in fig.~\ref{fig:PCAsymGeoVar} for the PC asymmetry.

\section{Commissioning and Chamber Performance}\label{scn:comm}

To ensure that the chamber performance satisfied the experimental goals, test were performed prior to installation, to establish its behavior under various bias voltage conditions and study noise behavior. Commissioning beam data was used to test linearity and noise behavior, measure possible false asymmetries associated with the electronics and measure the correlations between the wires, due to signal sharing as a result of the long proton path length, as well as to measure the correlation slopes between wire and beam monitor intensity, which were used in a regression analysis, to investigate possible false asymmetries due to beam fluctuations.

\subsection{Target Pressure and Breakdown Voltage Tests}\label{sec:n3HeVoltageScan}

The initial high voltage testing of the chamber in room air was successful at $600\,\mathrm{V}$ applied to the HV.  Further testing on the sealed, evacuated chamber produced a breakdown voltage of $2500\,\mathrm{V}$. To test the chamber near operating conditions, the breakdown voltage at two ${^4}$He fill gas pressures was measured. At $0.88\,\mathrm{atm}$ pressure sparking occurred at $940\,\mathrm{V}$, and at $0.41\,\mathrm{atm}$ sparking occurred at $800\,\mathrm{V}$.

\subsection{Bias Voltage Scan}
The charge collection efficiency of an ionization chamber depends on the applied bias voltage. To determine suitable operating conditions of the n$^3$He target chamber, beam-on voltage scans were done over an extended range, from $0$ to $-552\,\mathrm{V}$ in 18 voltages steps.
The quantity of interest for the charge collection efficiency test is the ratio of wire to beam monitor yields:
\begin{equation}\label{eqn:VoltScanRatio}
R_{Y_{i,n}} = \frac{\langle Y_{i,n}\rangle - \langle Y_{i,0}\rangle}{\langle M_n \rangle - \langle M_0 \rangle}~.
\end{equation}
Where $Y_{i,n}$ is signal for wire $i$ at bias step $n$, $M_n$ is the beam monitor signal for bias step $n$, while $Y_{i,0}$ and $M_0$ are the corresponding pedestals. For each of the 18 bias voltage steps a data run was taken over 2000 neutron pulses. To remove effects due to beam instability, neutron pulses for which the peak flux in the beam monitor varied by more than 10\% from the average were cut. Wire and beam monitor signal pedestals were determined by taking beam-off runs (with the neutron beam shutter closed).

\begin{figure}[h]
 \centering
  \includegraphics[width=0.36\textwidth]{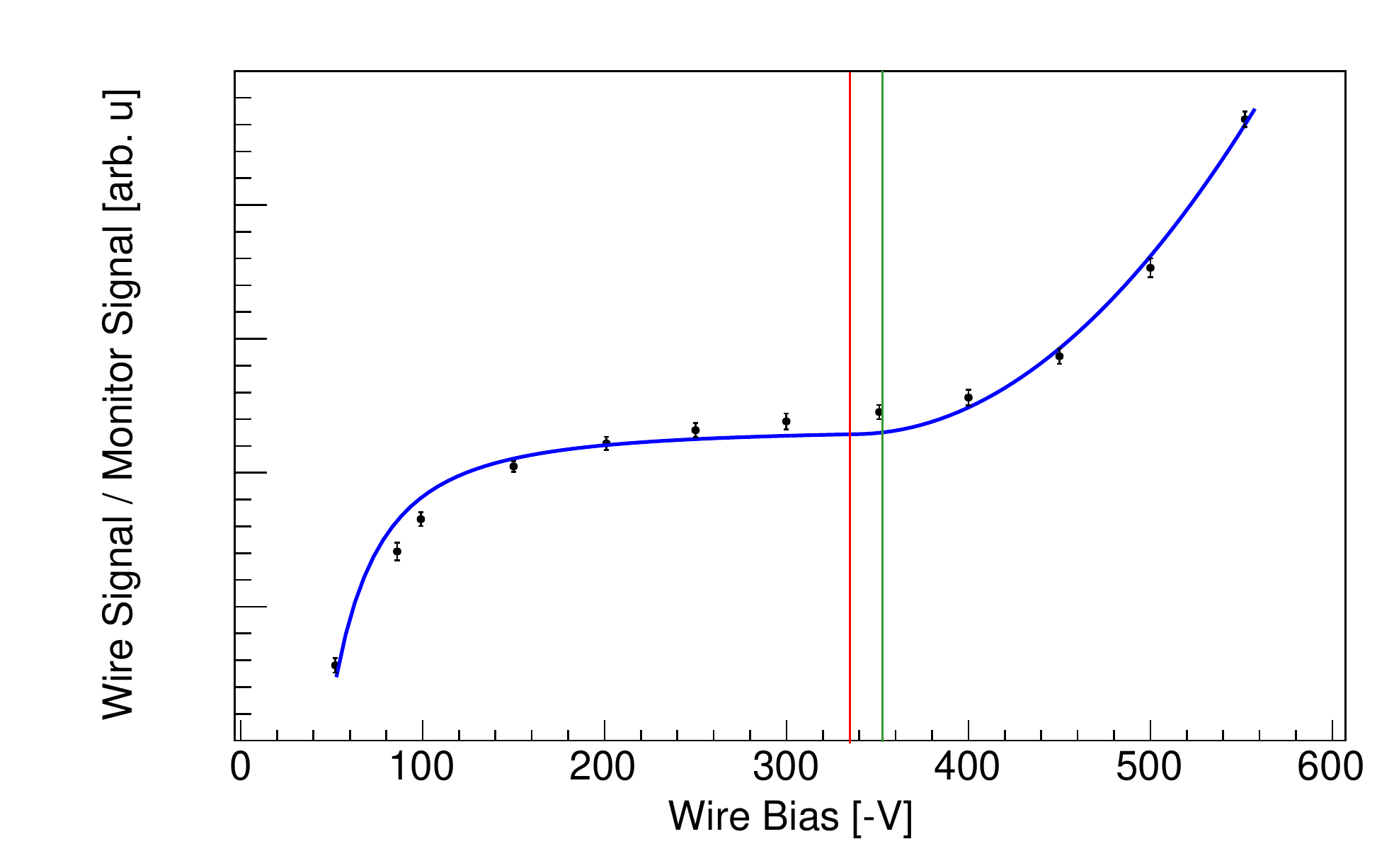}
 \caption{Wire signal divided by the beam monitor signal (both integrated over the TOF range), plotted versus the absolute value of the applied bias voltage. The red line shows the approximate onset of the amplification region at about $-340\;\mathrm{V}$. The operating voltage was selected to be $-353\;\mathrm{V}$ (the green line). The blue line illustrates the trend.}
 \label{fig:WirePulses64:Fit}
\end{figure}

Fig.~\ref{fig:WirePulses64:Fit} shows a typical voltage scan for one wire. The recombination region, where only a fraction of the initial ionization is collected, occurs up to approximately $-200\,\mathrm{V}$.  The voltage region for which the yield is flat with respect to bias voltage (the so called  ``ion chamber'' (IC) region) indicates full charge collection without amplification. Amplification begins around $-340\,\mathrm{V}$. The graph shows a qualitative fit, using
\begin{equation}\label{eqn:TwoPart}
  f(V_b) = I - \frac{C}{V_{t}^2} + \theta(V_b-V_{t})(V_b-V_{t})^2 I A ~.
\end{equation}
%\begin{align}
%  f(V_b) = \left\lbrace\begin{array}{cl}
%     I - \frac{C}{V_b^2} & \mathrm{if\ } V_b<V_{t}\\
%     I - \frac{C}{V_{t}^2} + I A (V_b-V_{t})^2 & \mathrm{if\ }V_b>V_{t}\\
%  \end{array}  \right. ~. \label{eqn:TwoPart}
%\end{align}
Where $I$ is the initial ionization, $C$ is the recombination factor, $V_b$ is the applied bias voltage, $V_{t}$ is the transition voltage from recombination to amplification. The signal from the outer most wires in each wire plane showed a different response to the high voltage than the inner wires on each plane, transitioning directly from recombination to amplification without showing an obvious IC region, while the inner wires showed a more extended IC signal region.  As can be seen in fig.~\ref{fig:wirelayout}, the charge collection region around the outermost wires is defined by two high voltage wires and the edge of the wire frames, while the inner wires are in surrounded by four high voltage wires.  The reduced number of field wires creates a different field configuration around these wires than the inner wires with four HV wires, which affected how charge collection occurs.

While the voltage for the transition from the IC to the amplification region was below the selected operating bias voltage of $-353\,\mathrm{V}$, the degree of amplification is small. Operation in the IC region would have been preferred, as it is less sensitive to power supply voltage variations, operating at the low end of the amplification region was deemed to be suitable. The primary concern is a buildup of excessive space charge at high ionization rates in the target chamber, which would reduce the chamber linearity.

\subsection{Target Linearity}\label{sec:n3HeLinearity}

The primary reason for establishing that the chamber was operating in or near the IC region was to make sure that the response of the wire signal yield to changes in ionization yield was linear. Since the asymmetry is the result of a systematic variation in the number of emitted protons in a given direction, with respect to the neutron spin, a non-linear increase of the wire yield with respect to the corresponding ionization would produce a scale error the asymmetry. The linearity tests were used to verify that the chamber operation is linear to within a few percent, which is    significantly better than the overall measurement precision of $\geq ~10\%$.

\begin{figure}[h]
 \centering
  \includegraphics[width=0.45\textwidth]{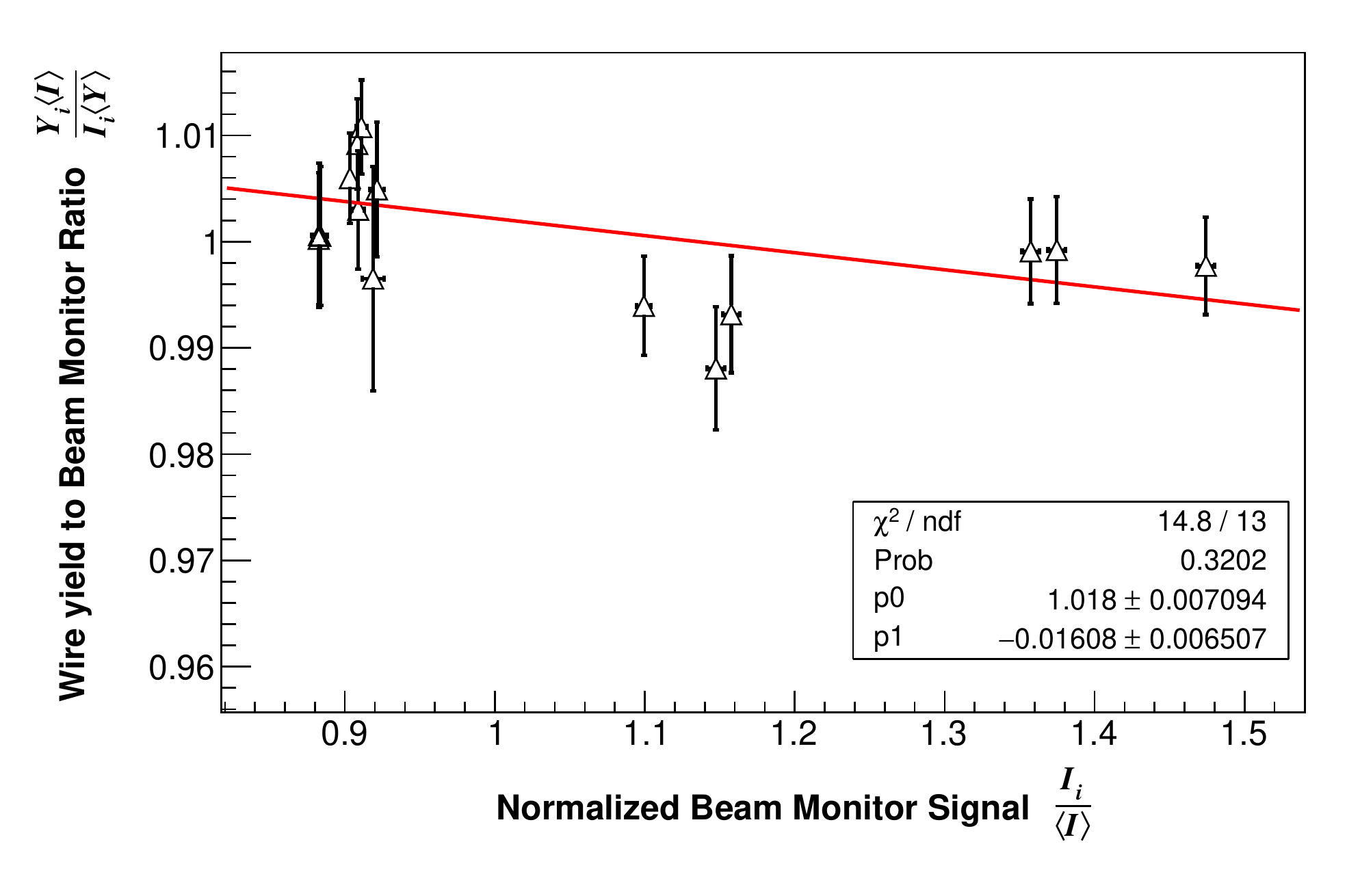}
 \caption{A fit to eqn.~\ref{eqn:nlin2} for one of the sense wires. The data and the fit show that non-linearities beyond first order are negligible. Extracting the non-linearity for each wire corresponds to calculating the ratio of the extracted slope and vertical axis intercept}
 \label{fig:OneWireLin}
\end{figure}

\begin{figure}[h]
 \centering
  \includegraphics[width=0.45\textwidth]{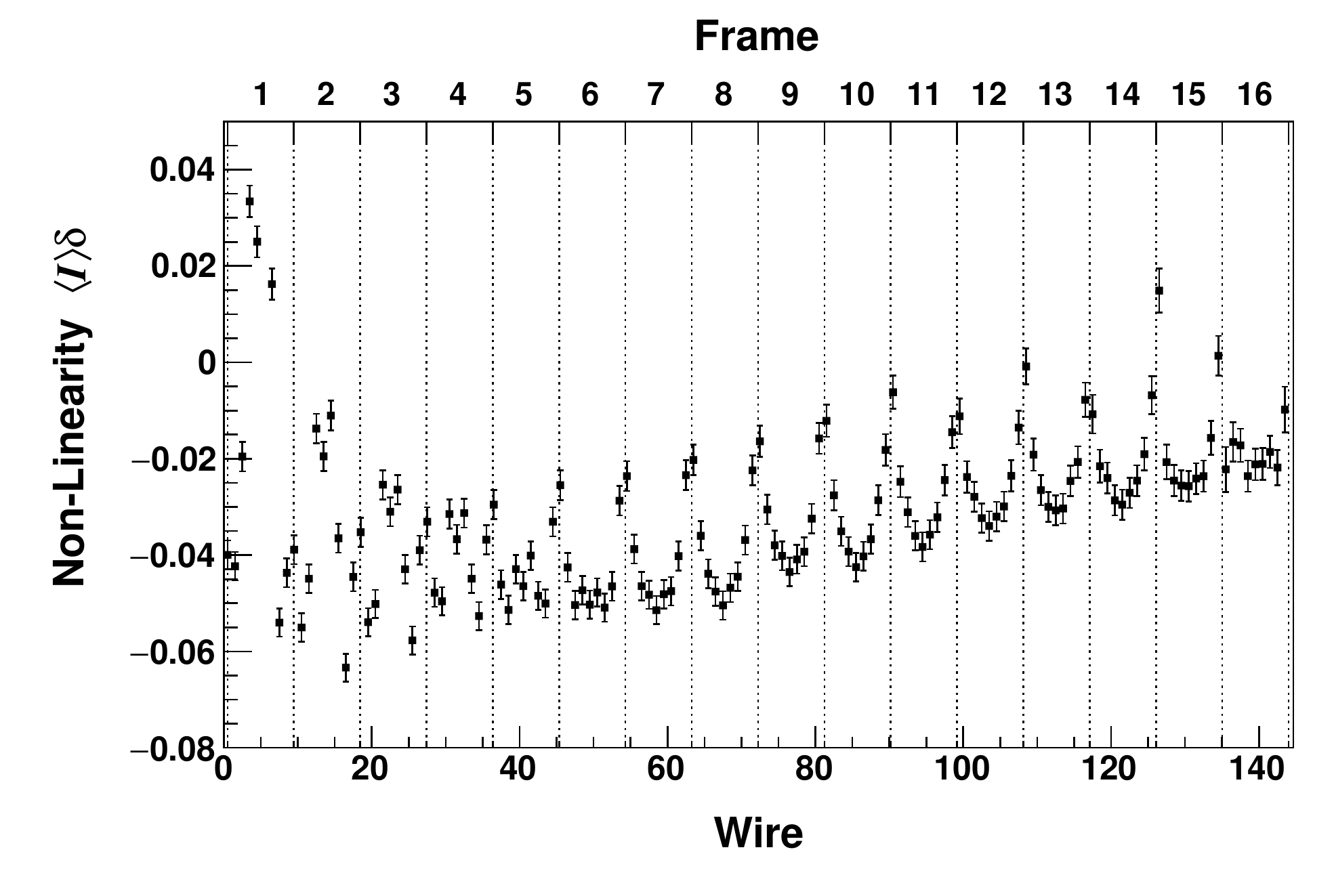}
 \caption{Non-linearity factors for all wires. The factors were extracted from a fit of the wire yield data as a function of beam intensity (see text).}
 \label{fig:AllWiresLin}
\end{figure}

Since it is difficult to setup a controlled current mode linearity measurement for ion chambers, a number of runs were selected, spread over the n$^3$He run time, such that the data covered the full range of the proton accelerator beam power from $850\,\mathrm{kW}$ to $1400\,\mathrm{kW}$, producing varying levels of capture intensity in the target. The corresponding variation in neutron beam intensity produces a change in ionization current in the chamber. Since the FnPB beam monitors were previously measured to be highly linear over a wide neutron flux range, one of them was used as a reference to measure the chamber linearity. Under the presence of a non-linearity, the signal yield on a wire, for run $i$ with a certain neutron flux, can be written as
\begin{equation}\label{eqn:nlin}
    Y_i = gI_i\left(1 + \delta f(I_i) \right) = gI_i\left(1 + \alpha I_i + \beta I^2_i + \ldots \right)~.
\end{equation}
Where $I_i$ is the neutron beam intensity measured with the beam monitors and $\delta$ sets the strength of the overall non-linearity in the wire signal, which is expressed as an arbitrary function of signal intensity $f(I_i)$, parameterized in terms of the beam monitor signal. To extract the non-linearity, one can plot the following unit-less ratio versus $I_i / \langle I \rangle$:
\begin{equation}\label{eqn:nlin2}
  R(I_i) = \frac{Y_i \langle I \rangle}{I_i \langle Y \rangle} = \frac{1 + \delta f(I_i)}{1+ \delta \frac{\langle I f(I) \rangle}{\langle I \rangle}}~.
\end{equation}
Where $\langle Y \rangle$ and $\langle I \rangle$ are the averages over all runs used in the linearity study. A typical example of $R(I_i)$
plotted against $I_i / \langle I \rangle$ is shown in Fig.~\ref{fig:OneWireLin}. The fit shows that non-linearities beyond first order are negligible and that one can replace $f(I_i)\rightarrow I_i$ in eqn.~\ref{eqn:nlin2}. The slope of the fit $\lambda = \langle I \rangle dR/dI_i$ and the zero intensity intercept $R(0)$ allows one to calculate the non-linearity $\lambda / R(0) = \langle I \rangle \delta $. The measured non-linearity for all wires is shown in Fig.~\ref{fig:AllWiresLin}.

With a non-linearity, the target chamber is more likely to lose efficiency at higher neutron flux, so that a negative slope is expected.  The reason for this is that as the ionization rate increases the space charge density in the gas volume also increases. Particularly the positive ions, due to their long collection times, can build up near the collecting electrodes.  This can result in charges in the target chamber being shielded from the HV, reducing the effective field, increasing collection times, and allowing for more recombination. The ionization current is largest at the center of the chamber, so that non-linearities are largest for the center wires, as evident in Fig.~\ref{fig:AllWiresLin}.

Equation~\ref{eqn:nlin} ignores asymmetries, because it was a study relying only on larger changes of wire signal yield, as a function of neutron beam intensity. Ultimately however, the non-linearity that was measured in this way was that of the response of the wire to changes in ionization current levels and the physics asymmetry itself corresponds to such a change in ionization current, sensed by a given wire. If one incorporates the asymmetry in the ionization current level $I$ , in a simplified expression of the form $I^{\pm} = I_o(1\pm A_{P})$ and the wire yield has a non-linear response to the current $Y^{\pm} = gI^{\pm}\left(1 + \delta I^{\pm} \right)$, then the measured asymmetry is given by
\begin{eqnarray}\label{eqn:nlin3} \nonumber
  A_{meas} = \frac{Y^+ - Y^-}{Y^+ + Y^-} &=& A_{P}\frac{1+2\delta I_o}{1+\delta I_o + \delta I_o A^2_{P}}  \\
                                         &\simeq & A_{P}\frac{1+2\lambda}{1+\lambda} ~.
\end{eqnarray}
For small slopes $A_{meas} \simeq (1+\lambda)A_{P}$. So as long as the measured slopes are at the few percent level, the non-linearity effect is small. Since the effect was measured for all wires, the correction was applied directly in the analysis.

\subsection{Detector Noise and Instrumental False Asymmetry}\label{sec:InstrumentalAsym}

Beam off data runs were taken throughout the running period of the experiment, whenever beam was off for maintenance or other reasons. This happened
regularly, at least every Tuesday during production data taking, and there were long beam off periods over the summer months. The beam-off data was used to study the detector noise behavior and establish the size of a possible false asymmetry associated with the experiment electronics. The latter is possible if the RFSR (radio-frequency spin rotator) produces a change in the detector electronics signal chain, due to load changes when the RFSR switches on and off. Direct, inductive coupling of the RFSR field into the electronics can also lead to false asymmetries.

\begin{figure}[h]
 \centering
  \includegraphics[width=0.45\textwidth]{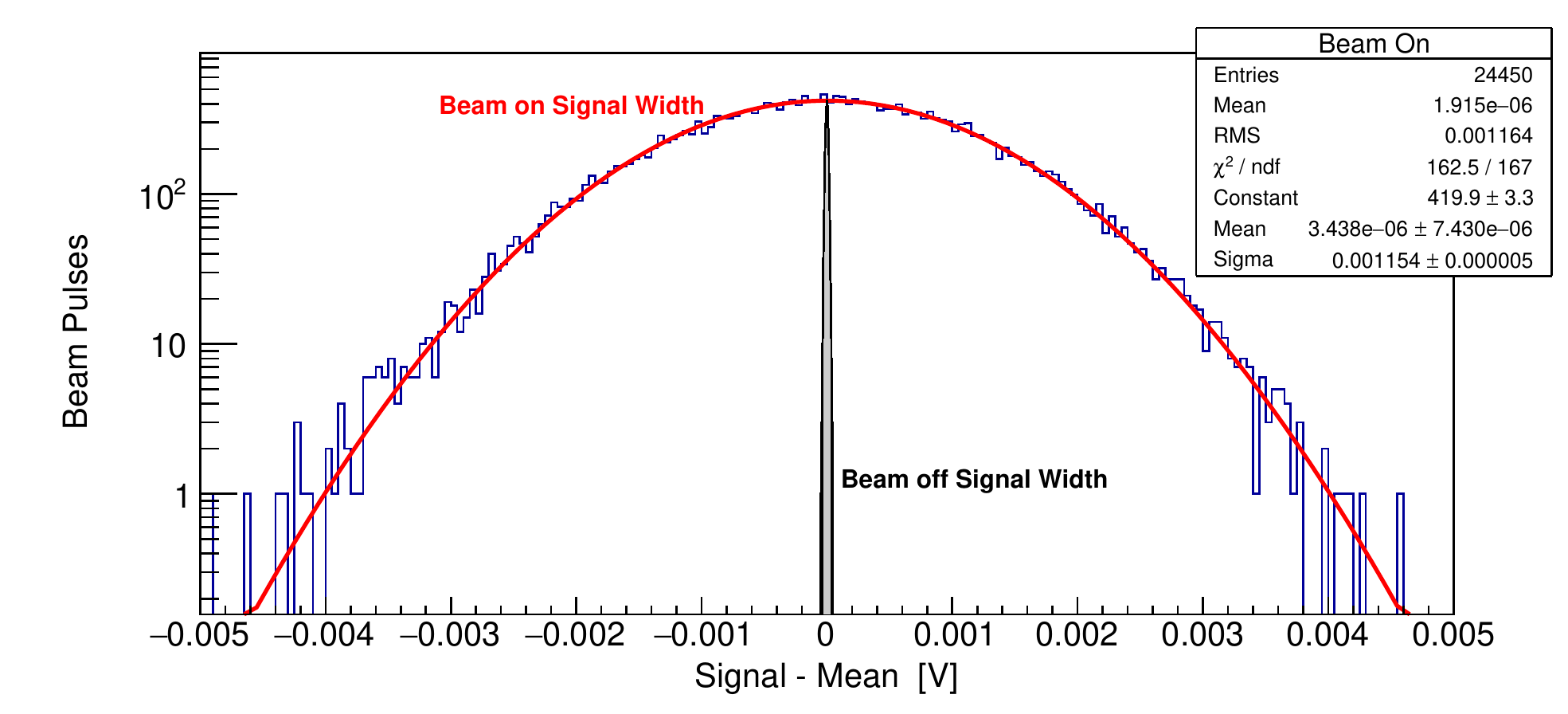}
 \caption{A comparison between the beam-on signal width (RMS) and the beam-off signal width (RMS) due to electronics noise. The horizontal axis corresponds to the signal deviation from the mean signal for the neutron energy with peak flux (see Fig.~\ref{fig:TwoPulsePreampOut}), within a group of 50 pulses with stable beam conditions. The much smaller RMS in the electronics ensures that any electronics false asymmetry can be determined with a small uncertainty, within a short run time.}
 \label{fig:RMSComp}
\end{figure}

The design of the experiment was chosen to minimize a possible electronics false asymmetry. Design measures included the implementation of isolated power circuits and avoiding ground loops and other ways for the fields to couple to the wire signal electronics and the data acquisition. To reduce the effect of any remaining false asymmetry, the most important aspect of the chamber and electronics performance is that the root-mean-square (RMS) electronics noise should be far below the beam-on width. The beam-on width is primarily due to neutron counting statistics, but is increased somewhat by the correlation between wires and the finite size neutron capture distribution. The relatively small size of the electronics signal RMS allowed the error on the false asymmetry to be determined to a level far below that of the statistical error, within a much shorter time period. As a result, any remaining false asymmetry can be applied as a correction, with negligible effect on the overall asymmetry error. A comparison between the electronics and counting statistics RMS can be seen in Fig.\ref{fig:RMSComp}. At the end of data taking the instrumental asymmetry was determined to be $A_{instr} = (0.3\pm2.0)\times 10^{-9}$.

\subsection{Beam Fluctuations and False Asymmetries}

\begin{figure}[h]
 \centering
  \includegraphics[width=0.45\textwidth]{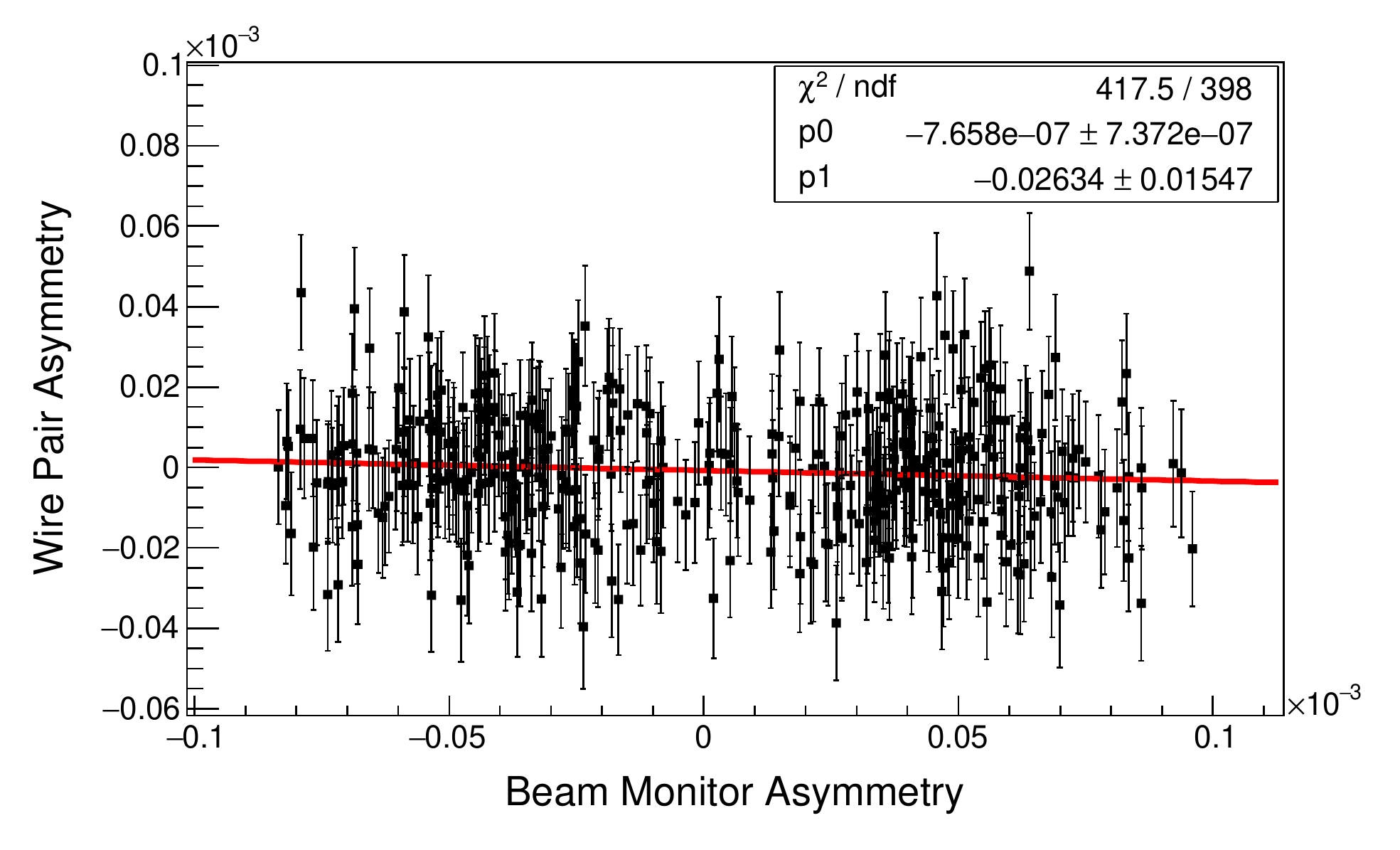}
 \caption{Linear Regression was used to extract the dependence of the measured wire asymmetry on a possible false asymmetry due to beam intensity fluctuations as measured with the beam monitors.}
 \label{fig:RegrSlope}
\end{figure}

\begin{figure}[h]
 \centering
  \includegraphics[width=0.45\textwidth]{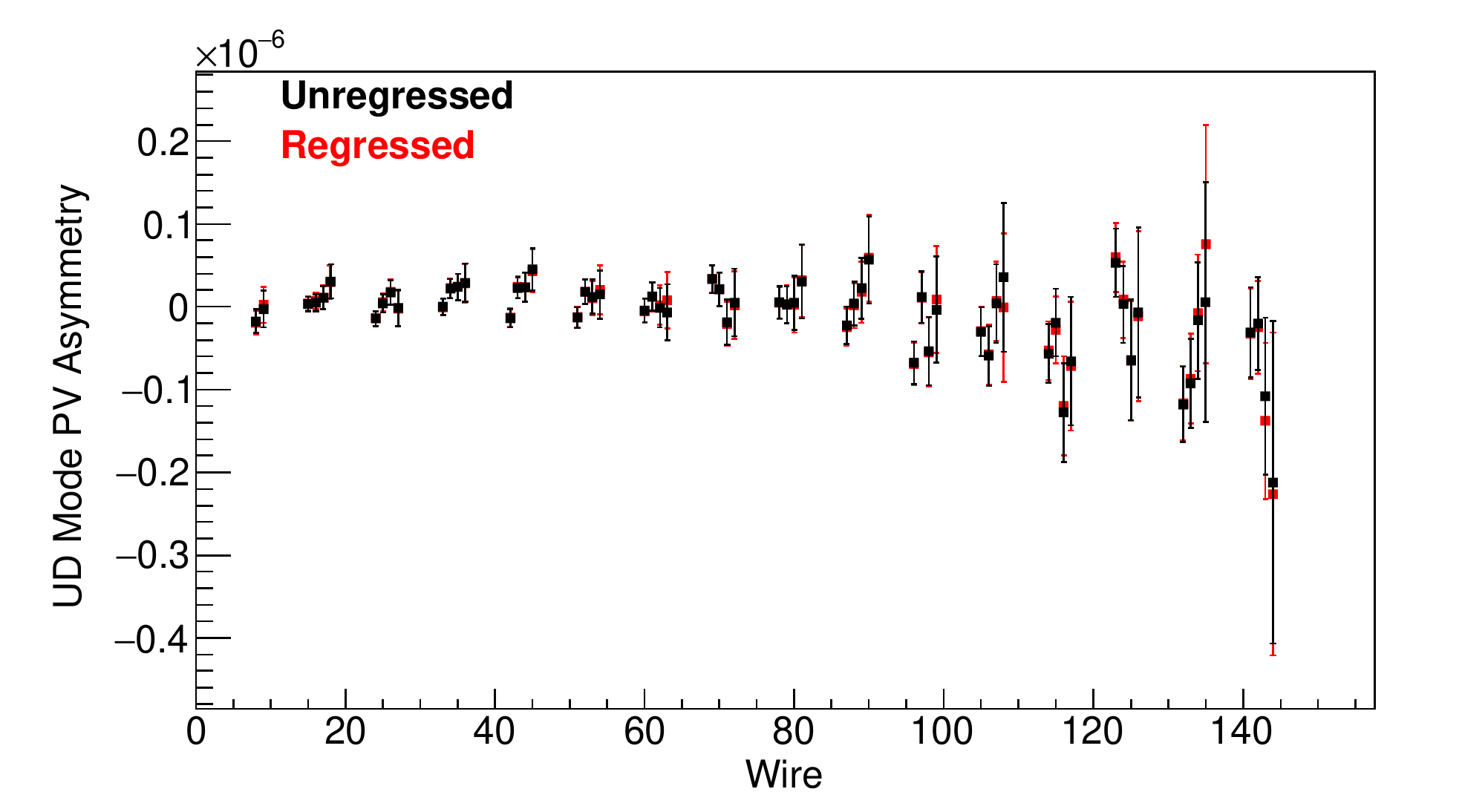}
 \caption{Comparison between the regressed and unregressed PV wire pair asymmetries, combined over the dataset. The small difference shows that the wire pair asymmetries did not have a strong dependence on the beam asymmetry.}
 \label{fig:AsymVsWire}
\end{figure}

The neutron beam can have large intensity fluctuations and can therefore produce false asymmetries. The beam monitors were used to measure the relative neutron beam intensity and pulse shape to $10^{-4}$ fractional uncertainty in the intensity for a single pulse. Regression analysis was used to investigate the effects of the beam fluctuations on the measured asymmetry. Equation~\ref{eqn:Asym} gives one of the two ways to form the wire pair asymmetry. To first order in the asymmetries, it is independent of the beam intensity asymmetry, but it depends on the pedestal asymmetry, which is measured in separate runs (see sec.~\ref{sec:InstrumentalAsym}). The other way to calculate the wire pair asymmetry is given by
\begin{eqnarray}\label{eq:Asy3} \nonumber
  A^{meas}_{i} &=& \frac{1}{2}\left(\frac{\frac{Y_{u,i}^{+}}{Y_{d,i}^{+}} - \frac{Y_{u,i}^{-}}{Y_{d,i}^{-}}}{\frac{Y_{u,i}^{+}}{Y_{d,i}^{+}} + \frac{Y_{u,i}^{-}}{Y_{d,i}^{-}}}\right)  \\ \nonumber
                &  & \\
               &\simeq & 2\epsilon P G^{PV}_{\mathrm{k}} A_{\mathrm{PV}} + b_i A_{Beam}~.
\end{eqnarray}
Where $b_i$ is the dimensionless slope of the plot $A^{meas}_{i}$ vs. $A_{Beam}$ which is extracted from the data using regression with respect to natural beam intensity modulation. An example of the extracted slope for one wire is shown in Fig~\ref{fig:RegrSlope}.
The full analysis process required two passes through the dataset, one to calculate the regression slopes and another to remove the beam asymmetry. Regressed asymmetries were then combined in error weighted averages to produce the final wire by wire asymmetries. Figure~\ref{fig:AsymVsWire} shows the difference between the regressed and unregressed PV wire pair asymmetries for the entire dataset. The regression slopes for most wires had magnitudes below $0.05$, fluctuating around zero and slightly increasing with depth into the target. The regression analysis changed the wire pair asymmetry central value by less than $0.04\times10^{-8}$.

\subsection{Wire Correlation and Final Asymmetry Calculation}\label{scn:CorrAna}

\begin{figure}[h]
 \centering
  \includegraphics[width=0.45\textwidth]{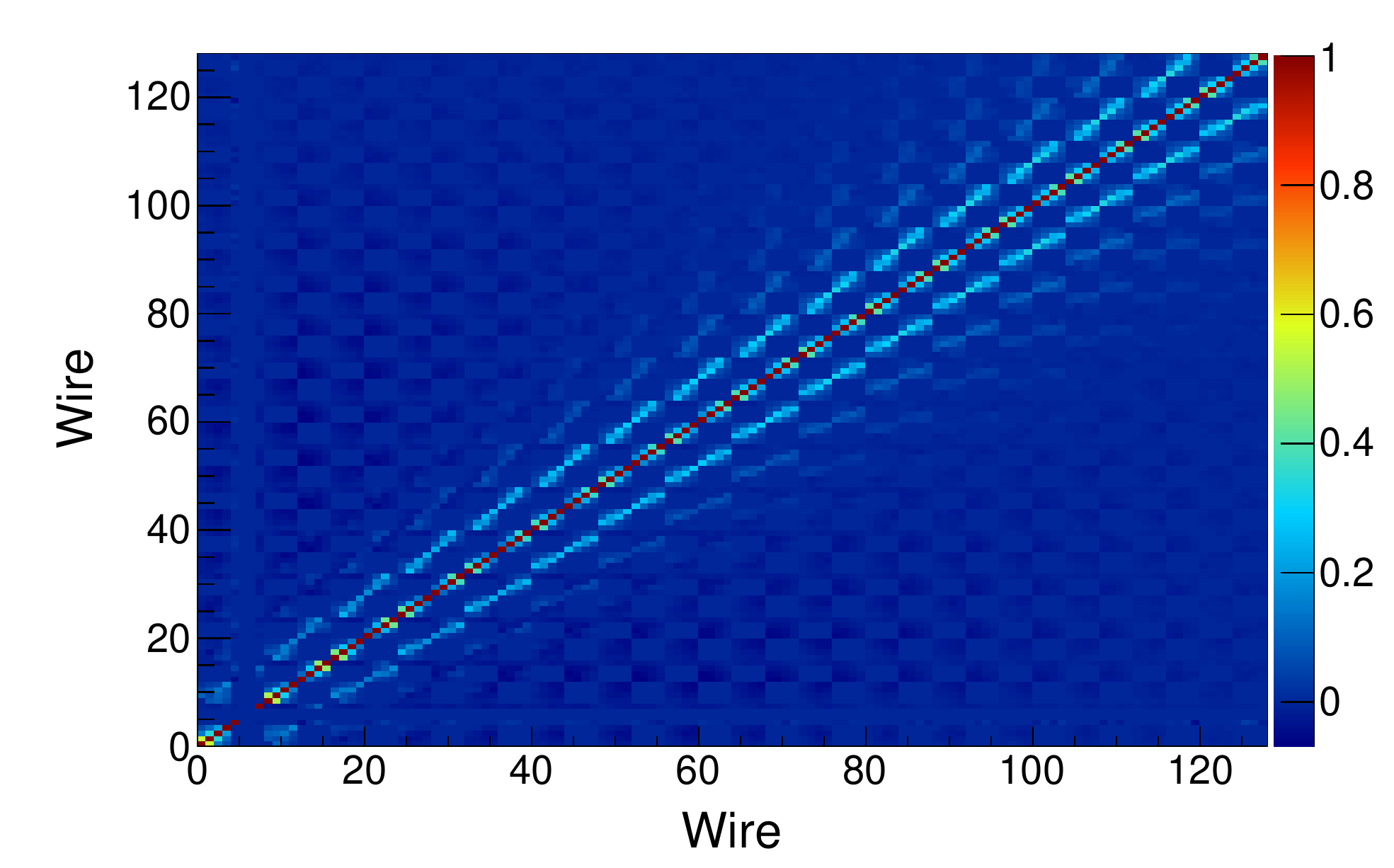}
 \caption{The measured correlation between wires.}
 \label{fig:WireCorr}
\end{figure}

Since the event rate did not allow for individual event detection, the long path length of the proton produced signal correlation among the wires, which needed to be taken into account in the data analysis. The correlation data was established from data and compared to simulations, where single capture event correlations could be studied. Figure~\ref{fig:WireCorr} shows the correlation for 128 wires, excluding the center wire in each frame.
The covariance matrix $(C_{ij})$ used in the analysis was formed from the the wire correlation and the errors obtained from the averages over the dataset for each wire $(i,j)$ (see Fig.~\ref{fig:AsymVsWire}). The final asymmetries were extracted by solving the matrix equation
\begin{equation}\label{eq:AsymMat}
    \vec{A} = \mathbf{M}^{-1}\vec{D}~.
\end{equation}
Where $\vec{A} = (A_{PV}, A_{PC})$, $\vec{D}$ is the data vector
\begin{equation} \nonumber
\vec{D} = \left(
\begin{matrix}
     G^{PV}_{UD,i} C^{-1}_{UD,ij} A^{meas}_{UD,i}  +  G^{PV}_{UD,i} C^{-1}_{UD,ij} A^{meas}_{LR,i} \\
     G^{PV}_{UD,i} C^{-1}_{UD,ij} A^{meas}_{UD,i}  +  G^{PV}_{UD,i} C^{-1}_{UD,ij} A^{meas}_{LR,i} \\
\end{matrix}
        \right)~,
\end{equation}
and $\mathbf{M}$ is the so-called \emph{measurement} matrix with elements
\begin{eqnarray}
\nonumber
  M_{1,1} &=& G^{PV}_{UD,i} C^{-1}_{UD,ij} G^{PV}_{UD,j} + G^{PV}_{LR,i} C^{-1}_{LR,ij} G^{PV}_{LR,j} \\ \nonumber
  M_{1,2} &=& G^{PV}_{UD,i} C^{-1}_{UD,ij} G^{PC}_{UD,j} + G^{PV}_{LR,i} C^{-1}_{LR,ij} G^{PC}_{LR,j} \\ \nonumber
  M_{2,1} &=& G^{PC}_{UD,i} C^{-1}_{UD,ij} G^{PV}_{UD,j} + G^{PC}_{LR,i} C^{-1}_{LR,ij} G^{PV}_{LR,j} \\ \nonumber
  M_{2,2} &=& G^{PC}_{UD,i} C^{-1}_{UD,ij} G^{PC}_{UD,j} + G^{PC}_{LR,i} C^{-1}_{LR,ij} G^{PC}_{LR,j} ~.
\end{eqnarray}
Equation~\ref{eq:AsymMat} is the result of a $\chi^2$ minimization with respect to $A_{PV}$ and $A_{PC}$. Note that this matrix would be diagonal without the frame twist described in Sec.~\ref{scn:geof}. To take into account the error from the geometry factor uncertainty, the minimization process was repeated many times, varying the geometry factors within their respective distributions shown in Figs.~\ref{fig:PVAsymGeoVar} and~\ref{fig:PCAsymGeoVar}.
Table~\ref{tbl:SYSEF} lists the uncertainties associated with the various effects discussed in this paper, as well as several other contributions that are discussed in other references. Taking all of these into consideration, the overall error on the asymmetry is $\pm 0.97~\mathrm{(Stat)} \pm 0.24~\mathrm{(Sys)})\times 10^{-8}$.

\begin{table*}%[H] add [H] placement to break table across pages
\caption{Systematic Corrections and Errors.\label{tbl:SYSEF}}
\begin{ruledtabular}
\begin{tabular}{cccc}
Additive Sources & Comment & Correction $[ppb]$ & Uncertainty $[ppb]$  \\ \hline
Frame Twist ($0$ to $20$ mrad) & compare simulation and data                 & $2.5$       & $0.2$   \\
Electronic false asymmetry     & measured                                    & $0.0$       & $2.0$   \\
Chamber field alignment        & compare simulation and data                 & $0.0$       & $1.3$\\
Mott-Schwinger scattering      & published calculation~\cite{MSCHW}          & $0.06$      & $0$ \\
Residual $^3$He Polarization   & calculation                                 & $< 0.06$    & $0$ \\
Background ($\beta$, $\gamma$) & simulation and calculation                  & $<< 0.1$    & $0$ \\
In-flight $\beta$-decay        & calculation~\cite{Blyth:2018aon}            & $<< 0.1$    & $0$ \\
Stern-Gerlach steering         & measurement and calculation ($\leq 2mG/cm$) & $<< 0.1$    & $0$ \\
\hline
Total                          &                                             & $2.6$       & $2.38$ \\
\hline \hline
Multiplicative Sources & comment & Correction & Uncertainty \\ \hline
Geometry factors               & compare simulation and data                 & $0.0$       & $0.5$ \\
Polarization                   & measurement~\cite{Hayes}                    & $0.936$     & $0.002$ \\
Spin-flip efficiency           & measurement~\cite{Hayes}                    & $0.998$     & $0.001$ \\
\hline
Total uncertainty              &                                             &             & $2.43$ \\
\end{tabular}
\end{ruledtabular}
\end{table*}

Data was collected over the period of a year, including production and commissioning runs. The number of processed (after beam cuts) parity-violating data runs was 31854, each of them 7 minutes long. The remaining runs included parity conserving data, beam studies, beam-off pedestal runs, as well as other commissioning related runs, including those that were used to benchmark the geometry factor and wire correlation simulations. To determine the geometry factors and verify the wire correlation extracted from beam data, well over $10^{10}$ neutron events were simulated with many systematic variations.

\section{Conclusion}
The n$^3$He experiment was constructed to make a precision measurement of the PV asymmetry in the $^{3}\mathrm{He}(\mathrm{n},\mathrm{p})^{3}\mathrm{H}$ reaction, with respect to the outgoing proton momentum. The measurement has the smallest uncertainty on any parity-violating observable made so far and provides a precision benchmark towards a complete determination of the weak hadronic coupling constants, both in traditional meson-exchange pictures, as well as effective field theories.

%%%%%%%%%%%%%%%%%%%%%%%%%%%%%%%%%%%%%%%%%%%%%%%%%%%%%%%%%%%%%%%%%%%%%%%%%%%%%%%
%   Acknowledgements                                                          %
%%%%%%%%%%%%%%%%%%%%%%%%%%%%%%%%%%%%%%%%%%%%%%%%%%%%%%%%%%%%%%%%%%%%%%%%%%%%%%%

\section*{Acknowledgments}
We gratefully acknowledge the support of the U.S. Department of Energy Office of Nuclear Physics through grant No. DE-FG02-03ER41258,
DE-AC05-00OR22725, DE-SC0008107 and DE-SC0014622, the US National Science Foundation award No: PHY-0855584, the Natural Sciences and Engineering Research Council of Canada (NSERC), and the Canadian Foundation for Innovation (CFI). This research used resources of the Spallation Neutron Source of Oak Ridge National Laboratory, a DOE Office of Science User Facility. We also thank Michele Viviani (INFN Pisa) for very fruitful theory discussions and Jack Thomison (ORNL) for his design support.

%%%%%%%%%%%%%%%%%%%%%%%%%%%%%%%%%%%%%%%%%%%%%%%%%%%%%%%%%%%%%%%%%%%%%%%%%%%%%%%
%   Bibliography                                                              %
%%%%%%%%%%%%%%%%%%%%%%%%%%%%%%%%%%%%%%%%%%%%%%%%%%%%%%%%%%%%%%%%%%%%%%%%%%%%%%%

\bibliographystyle{elsarticle-num}
\bibliography{n3HeChamber}
%bilbliography is currently dropping the year if pages are not given,

\end{document}